\documentclass[11pt]{iopart}

\usepackage{graphicx}
\usepackage{amssymb}
\usepackage{float}
\usepackage{tabularx}
\usepackage{microtype}

\usepackage[hidelinks,pagebackref=false,colorlinks=true,allcolors=blue]{hyperref}	% Allows for embedded clickable and back reference at the page where the reference was used
\begin{document}

%\begin{frontmatter}


\title[MAST-U spectroscopy detachment]{The role of plasma-atom and molecule interactions on power \& particle balance during detachment on the MAST Upgrade Super-X divertor}



\author{K. Verhaegh$^{1}$, B. Lipschultz$^2$, J.R. Harrison$^1$, F. Federici$^{2,8}$, D. Moulton$^1$, N. Lonigro$^{2,1}$, S. P. Kobussen$^5$, M. O'Mullane$^{3,1}$ N. Osborne$^4$, P. Ryan$^1$, T. Wijkamp$^{5,6}$, B. Kool$^{6,5}$, E. Rose$^1$, C. Theiler$^7$, A. Thornton$^1$ and the MAST Upgrade team$^*$}

\address{$^1$ United Kingdom Atomic Energy Agency, Culham, United Kingdom} 
  \address{$^2$ York Plasma Institute, University of York, United Kingdom}
\address{$^3$ University of Strathclyde, Glasglow, United Kingdom}
\address{$^4$ University of Liverpool, Liverpool, United Kingdom}
\address{$^5$ Eindhoven University of Technology, Eindhoven, The Netherlands}
\address{$^6$ DIFFER, Eindhoven, The Netherlands}
\address{$^7$ Swiss Plasma Centre, \'{E}cole Polytechnique F\'{e}d\'{e}rale de Lausanne, Lausanne, Switzerland}
\address{$^8$ Oak Ridge National Laboratory, Oak Ridge, Tennessee, USA}
\address{$^*$ See author list of “J. Harrison et al 2019 Nucl. Fusion 59 112011 (https://doi.org/10.1088/1741-4326/ab121c)“}

%\author[CCFE,York,EPFL}{Kevin Verhaegh}
%\author[York]{Bruce Lipschultz}
%\author[EPFL]{Basil Duval}
%\author[York,CCFE]{Alexandre Fil}
%\author[EPFL]{Olivier F\'{e}vrier}
%\author[SUPA,CCFE]{Daljeet Singh Gahle}
%\author[CCFE]{James Harrison}
%\author[CCFE]{David Moulton}
%\author[DIFFER]{Artur Perek}
%\author[EPFL]{Christian Theiler}
%\author[EPFL]{Mirko Wensing}
%\author[York]{Chris Bowman}
%\author[York]{Fabio Federici}
%\author[York,CCFE]{Omkar Myatra}

%\address[York]{York Plasma Institute, University of York, United Kingdom}
%\address[CCFE]{Culham Centre for Fusion Energy, Culham, United Kingdom}
%\address[EPFL]{Swiss Plasma Centre, \'{E}cole Polytechnique F\'{e}d\'{e}rale de Lausanne, Lausanne, Switzerland}
%\address[SUPA]{SUPA, University of Strathclyde, Glasgow, United Kingdom}
%\address[DIFFER]{DIFFER, Eindhoven, The Netherlands}

\ead{kevin.verhaegh@ukaea.uk}

\begin{abstract}
This paper shows first quantitative analysis of the detachment processes in the MAST Upgrade Super-X divertor (SXD). We identify an unprecedented impact of plasma-molecular interactions involving molecular ions (likely $D_2^+$), resulting in strong ion sinks (Molecular Activated Recombination - MAR), leading to a reduction of ion target flux. The MAR ion sinks exceed the divertor ion sources before electron-ion recombination (EIR) starts to occur, suggesting that significant ionisation occurs outside of the divertor chamber. In the EIR region, $T_e \ll 0.2$ eV is observed and MAR remains significant in these deep detached phases. The total ion sink strength demonstrates the capability for particle (ion) exhaust in the Super-X Configuration. 

Molecular Activated Dissociation (MAD) is the dominant volumetric neutral atom creation process can lead to an electron cooling of 20\% of $P_{SOL}$. The measured total radiative power losses \emph{in the divertor chamber} are consistent with inferred hydrogenic radiative power losses. This suggests that intrinsic divertor impurity radiation, despite the carbon walls, is minor in the divertor chamber. This contrasts previous TCV results, which may be associated with enhanced plasma-neutral interactions and reduced chemical erosion in the detached, tightly baffled SXD. 

The above observations have also been observed in higher heat flux (narrower SOL width) type I ELMy H-mode discharges. This provides evidence that the characterisation in this paper may be general. 
\end{abstract}

%\vspace{1pc}

\section{Introduction}
\label{ch:introduction}

The successful development of fusion energy faces a significant obstacle in the challenge of divertor power exhaust, as the heat flux directed at the target must be significantly reduced to meet engineering limits \cite{Pitts2019,Wenninger2014}. The plasma target heat flux $q_t$ is described by equation \ref{eq:plasmaheat} where $\gamma$ is the sheath transmission factor, $T_t$ is the target temperature, $\epsilon$ is the surface recombination energy 13.6 eV and $\Gamma_t$ (ions/s/$m^2$) is the target ion flux \footnote{Whilst dissociating a molecule $D_2$ costs 4.4 eV of energy.}.

\begin{equation}
    q_t = \Gamma_t  (\gamma T_t + \epsilon)
    \label{eq:plasmaheat}
\end{equation}

To reduce $q_t$ sufficiently in reactors plasma detachment is required, which occurs when plasma-neutral interactions result in simultaneous power, momentum, and particle (i.e., ion) losses, effectively decreasing the ion target flux \cite{Verhaegh2021b,Lipschultz1999,Stangeby2018,Krasheninnikov2017} whilst the target temperature is kept constant or is decreasing. Detachment occurs when the divertor target plasma temperature is reduced to below approximately 5 eV, which can be achieved by increasing the core density or introducing extrinsic impurity seeding to induce radiative power losses (a necessary requirement for reactors). 

Alternative divertor configurations (ADCs) are being developed that are predicted to tackle the power exhaust challenge by leveraging variations in divertor magnetic topology and enhanced neutral baffling. This may serve as a risk mitigation strategy if conventional divertors in reactors cannot withstand the power exhaust challenge. One example of an ADC is the tightly baffled Super-X divertor, used by the novel MAST Upgrade tokamak. With the Super-X, the strike point is shifted to a larger major radius, leading to a larger gradient in the magnetic field along the flux tubes to the target, which reduces the heat flux and plasma temperature and makes plasma detachment more accessible \cite{Havlickova2015,Theiler2017,Moulton2023,Verhaegh2023}.

\subsection{Detachment physics}

Particle balance (equation \ref{eq:ParticleBalance}) implies that the integrated ion target flux ($I_t=\int \Gamma_t dA$ in ions/s) is equal to the divertor ion source ($I_i$) minus the divertor ion sinks ($I_r$) plus any net influx from ions outside of the divertor towards the target ($I_u$). 

\begin{equation}
I_t = I_i - I_r + I_u
\label{eq:ParticleBalance}
\end{equation}

Generally, the divertor ion target flux is much higher than any flow of ions from upstream ($I_t \gg I_u$), which implies that the divertor ion source dominates any upstream flows ($I_i \gg I_u$) \footnote{However, this is not necessarily the case for the MAST Upgrade Super-X divertor as will be shown in this work (figure \ref{fig:PartBal})}. In these conditions, the divertor is in 'high recycling' conditions and the 'closed box' approximation ($I_t \approx I_i - I_r$) is valid. 

Since it takes $E_{ion}$\footnote{The ionisation energy $E_{ion}$ represents the potential energy required to turn a molecule into an ion ($\epsilon$) plus the radiative energy lost due to electron-impact excitation preceding ionisation \cite{Verhaegh2019}.} of energy to ionise hydrogen, power and particle balance are intertwined. Under the closed box approximation, this leads to the ion target flux estimate provided by equation \ref{eq:PowerPartBal}. This means that reducing the ion target flux (if $T_t \ll \frac{E_{ion}}{\gamma} \rightarrow T_t \ll 4-6 eV$, such that the second term of equation \ref{eq:PowerPartBal} is roughly 1) requires either ion sinks ($I_r$) and/or a reduction of $\frac{P_{recl}}{E_{ion}}$. The latter leads to power limitation \cite{Verhaegh2019} (or "starvation" \cite{Lipschultz1999,Krasheninnikov1997}) of the ionisation source, which can be achieved by reducing $P_{recl}$ (e.g., the power entering the recycling region) through impurity seeding. 

\begin{equation}
I_t = (\frac{P_{recl}}{E_{ion}} - I_r) \times \frac{1}{1 + \frac{\gamma T_t}{E_{ion}}}
\label{eq:PowerPartBal}
\end{equation}

However, due to the marginal Bohm criterion at the plasma sheath ($\Gamma_t \propto p_t / \sqrt{T_t}$) \footnote{The Bohm criteria applies to a single flux tube, but for simplicity below we will apply it to the integrated ion target flux profile}, any reduction of the ion target flux requires the target pressure ($p_t$) to drop faster than the square root of the target temperature ($\sqrt{T_t}$). Such target pressure losses can be brought on by either upstream pressure losses \cite{Krasheninnikov2017,Kukushkin2017}, which are undesirable for a reactor but sometimes observed experimentally \cite{Verhaegh2019,Fevrier2019submitted}; or volumetric momentum losses \cite{Stangeby2018,Stangeby2017} which reduce the pressure before reaching the target. Taking the marginal Bohm criterion and momentum balance into account, the ion target flux can be modelled using equation \ref{eq:MomBalance}.

\begin{equation}
I_t = \frac{\gamma p_t^2}{2 m_i P_{recl}} \frac{\frac{\gamma T_t}{E_{ion}}}{1 + \frac{\gamma T_t}{E_{ion}}}
\label{eq:MomBalance}
\end{equation}

Although equations \ref{eq:MomBalance} and \ref{eq:PowerPartBal} are different approaches (e.g. one focuses on momentum balance, whereas the other focuses upon power/particle balance), it has been shown that both formulations are equivalent \cite{Verhaegh2019a}. Any reactor-relevant detached solution will require simultaneous power, momentum and particle losses.

\subsection{Plasma detachment and spectroscopy}

Plasma detachment is often studied by observing the effects of detachment on various parameters, such as reduced heat flux to the target ($q_t$), decreased ion target flux ($I_t$), and increased radiative losses. However, this study aims to explain the microscopic origin of these macroscopic results by examining plasma-atom/molecular interactions. In the 1990s, research showed that electron-ion recombination can be a crucial ion sink (contributing to $I_r$) during detachment under certain conditions \cite{Terry1998,Terry1999,Lipschultz1999,Verhaegh2017}. This could explain the ion target flux ($I_t$) roll-over in some, but not all, cases, which led to the suspicion that the ion target flux reduction may be due to a decrease in the ionisation source ($I_i$), caused by power limitation. Here, the power entering the recycling region ($P_{recl}$) becomes comparable to the power required for ionisation ($I_i E_{ion}$) \cite{Lipschultz1999,Krasheninnikov1997,Verhaegh2019,Verhaegh2018}. Later research confirmed the reduction of the ionisation source during detachment on JET \cite{Lomanowski2019} \& TCV \cite{Verhaegh2019}. TCV results showed that the reduction of the divertor ionisation source ($I_i$) was indeed correlated to power limitation \cite{Verhaegh2019}.

However, this purely atomic analysis \cite{Lomanowski2019,Verhaegh2019,Verhaegh2019a} could not explain all spectroscopic observations, particularly the brightness of the $D\alpha$ emission. Further research on JET \cite{Karhunen2022}, TCV \cite{Verhaegh2021,Verhaegh2021a,Verhaegh2021b} and MAST-U \cite{Verhaegh2023} revealed that this emission arises from excited atoms born from plasma-molecular interactions involving molecular ions ($D_2^+$ and/or $D_2^- \rightarrow D^- + D$). When those ions react with the plasma, they result in additional hydrogenic emission \& power losses (hydrogenic radiation, potential energy losses), ion sources (Molecular Activated Ionisation - MAI), ion sinks (Molecular Activated Recombination - MAR - contributing to $I_r$), and neutral atom sources (Molecular Activated Dissociation - MAD). Analysis has shown that these interactions can lead to strong ion sinks that are more significant than electron-ion recombination on TCV and play a crucial role in the ion target flux reduction \cite{Verhaegh2021a, Verhaegh2021b}. This increased the total ion sink ($I_r$) significantly compared to older studies that did not include plasma-molecular interactions \cite{Verhaegh2019}. With the increase in the total ion sink, the ion target flux was significantly larger than the divertor ion sources minus divertor ion sinks during deep detachment ($I_t \ll I_i - I_r$ \cite{Verhaegh2021b}. This suggests the ion flow from outside the divertor towards the target ($I_u$) is significant during deep detachment and high recycling conditions stop applying. This is in agreement with SOLPS-ITER modelling predictions \cite{Verhaegh2019,Fil2017} as well as data from a novel divertor scanning reciprocating probe \cite{Oliveira2022}.

The finding that plasma-molecular chemistry plays a major role during detachment on TCV was inconsistent with plasma-edge simulations, which showed negligible amounts of molecular activated recombination and dissociation from molecular ions and associated $D\alpha$ emission, as well as a lack of ion target flux roll-over during detachment \cite{Verhaegh2021b,Fil2017,Wensing2019}. This was attributed to underestimation of molecular charge exchange in Eirene, particularly for deuterium and tritium \cite{Verhaegh2021b,Verhaegh2023a,Reiter2018,Kukushkin2017}. Increasing the molecular charge exchange ($D_2 + D^+ \rightarrow D_2^+ + D$) cross-section by disabling ion isotope mass re-scaling from H to D, both through post-processing the TCV SOLPS-ITER simulations \cite{Verhaegh2021b}, as well as self-consistent SOLPS-ITER simulations \cite{Williams2022,Verhaegh2023a}, led to an improved match between the experiment and simulation.  

\subsection{Detachment in the MAST Upgrade Super-X divertor}

The results from the first MAST Upgrade campaign indicate both a further reduction of the target heat fluxes as well as a reduced detachment onset, in terms of the core density in the Super-X Divertor (SXD) compared to the Conventional Divertor (CD), required for detachment. These observations are roughly consistent with analytic predictions and simulations \cite{Moulton2023,Verhaegh2023,Havlickova2015,Moulton2017}. Divertor spectroscopy analysis of density ramp discharges in the Super-X divertor configuration was performed to separate the $D\alpha$ emission in terms of its different atomic and molecular processes. This indicated that, in a fuelling scan, detachment starts with the electron-impact excitation emission detaching from the target \cite{Verhaegh2023,Wijkamp2023}. This results in a region with a high molecular density below the ionisation region. Collisions, reactions and plasma-wall interactions can excite molecules vibrationally, which facilitates the creation of molecular ions (e.g. $D_2^+$ and/or $D_2^- \rightarrow D^- + D$). Those ions react with the plasma, leading to strong hydrogen Balmer line emission from the resultant excited neutral atoms \cite{Verhaegh2023,Wijkamp2023}. These observations are qualitatively consistent with previous TCV observations \cite{Verhaegh2021b,Verhaegh2021}, which were conducted in a single null conventional (open) divertor geometry before baffles were installed on TCV. However, plasma-molecular effects have a stronger impact on the hydrogenic emission in MAST-U.

\begin{figure}
\centering
\includegraphics[width=0.7\linewidth]{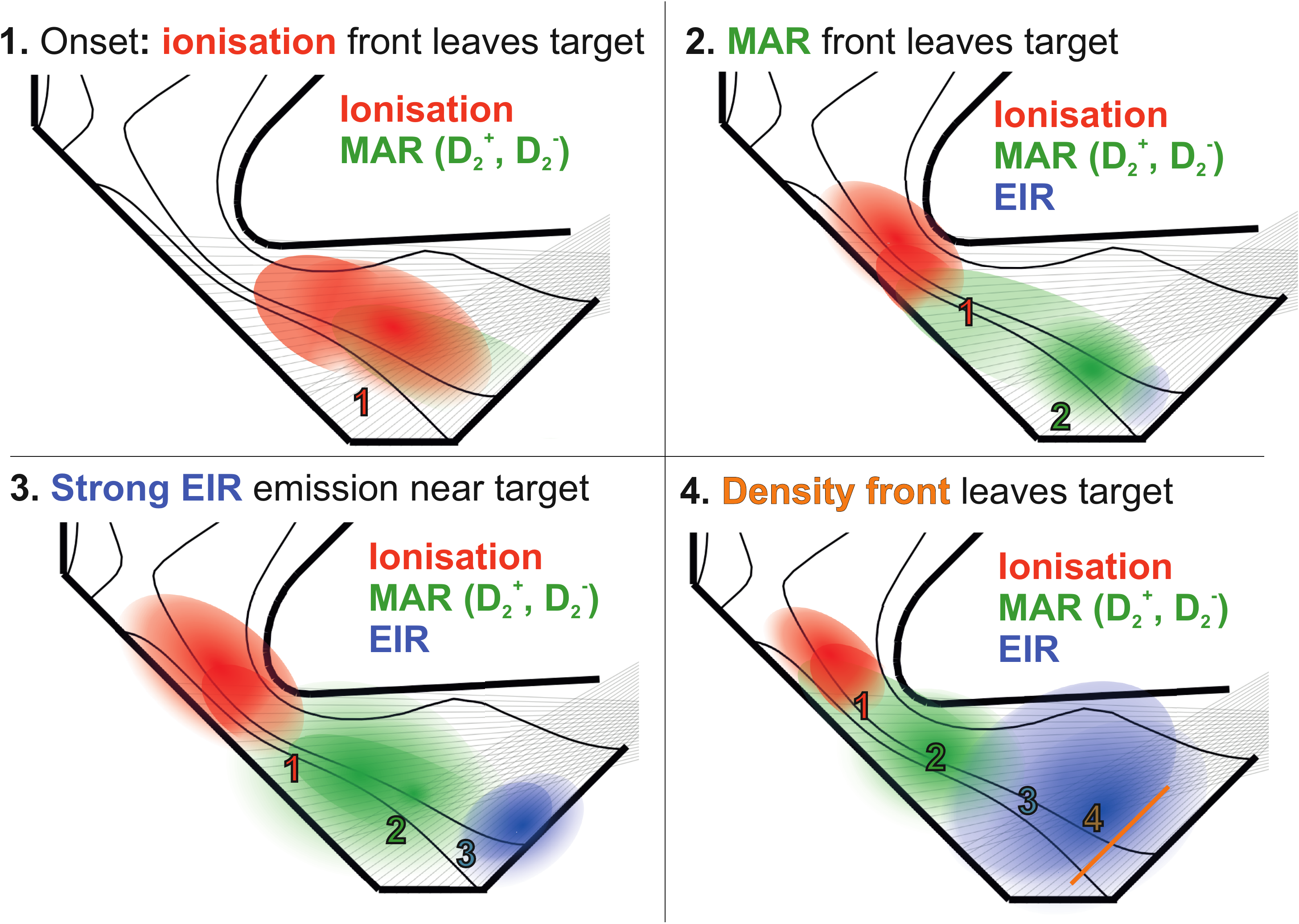}
\caption{Schematic overview of the four inferred MAST-U Super-X detachment phases in terms of the reactions occurring in the divertor. Also shown is the Super-X plasma geometry and the DMS spectroscopic viewing chords. The numbers shown indicate the detachment of the following regions from the target: (1) the back-end of the ionisation region; (2) the back-end of the Molecular Activated Recombination (MAR) region; (3) the front-end of the Electron Ion Recombination (EIR) region; and (4) the back-end of the electron ion recombination / density region.  The magnetic geometry in this illustration has been obtained from a SOLPS-ITER simulation (from \cite{Myatra2023,Myatra2023a}). Adopted from \cite{Verhaegh2023}.}
\label{fig:DetachmentPhases}
\end{figure}

Detachment in the Super-X divertor configuration can be distinguished by four phases \cite{Verhaegh2023,Wijkamp2023}, which are listed below and illustrated schematically in figure \ref{fig:DetachmentPhases}, adopted from \cite{Verhaegh2023}. 

\begin{enumerate}
\item The ionisation region detaches from the target. The new region between the ionization region and the target is characterised by strong Balmer emission from plasma-molecular interactions including MAD and MAR. 
\item The peak in Balmer line emission from plasma-molecular interactions detaches from the target as the divertor temperature drops below 1 eV and the efficiency of creating molecular ions is reduced.
\item Signs of electron-ion recombination (EIR) start to appear near the target, resulting in an increase of the higher/lower -n Balmer line ratio towards the EIR limit as well as high-n ($n\geq 9$) Balmer line emission. Temperature estimates of $\leq 0.2$ eV are found. 
\item The peak in EIR emission detaches from the target, which is consistent with a strong reduction of the electron density near the target \cite{Verhaegh2023}. 
\end{enumerate}

\subsection{This paper}

Building on the qualitative work in \cite{Verhaegh2023}, in this work we present a first quantitative analysis of the divertor ion sources \& sinks in the novel MAST-U Super-X divertor. This was enabled using new ADAS data \cite{OMullane} for EIR which has been extended to go below 0.2 eV. Our results indicate a strong or dominant presence of MAR ion sinks in the entire divertor chamber from detachment phase I until phase IV. In the deepest detached phases, where EIR ion sinks are significant and $T_e < 0.2$ eV is reached, MAR ion sinks remain important. The ion sinks are significantly stronger than the divertor ion source for the majority of the detached regime, which has implications for the plasma flow profile. Deep detachment states where ion sinks in the divertor chamber are similar or larger than the divertor ion source are found both during Ohmic L-mode as well as ELM-free and type-I ELMy Ohmic H-mode operation.

Extrapolating our Balmer line analysis to the total hydrogenic radiative losses and comparing this with the total radiation (i.e., separate divertor bolometry measurements) suggests that the total radiative power in the divertor is dominated by atomic hydrogenic radiation. Hydrogenic power losses can remain significant even when the ionisation source has moved upstream out of the divertor chamber due to power losses associated with plasma-molecular interactions, particularly MAD which, by then, is the dominant volumetric neutral atom generation process in the plasma.

\section{MAST-U overview}
\label{ch:MASTU_overview}

First, we will briefly discuss the three discharges used in this work. These discharges have no external heating (i.e., only Ohmic heating) and are already detached when the Super-X configuration is formed. Table \ref{tab:Overview} shows some of the general discharge parameters.

\begin{table}[]
\begin{tabular}{lllllll}
Discharge & $I_p$ (kA) & $P_{SOL}$ (kW) & $dr_{sep}$ (mm) & $f_{GW}$(\%) & Fuelling       & Description          \\
45371     & 650        & 470            & 1               & 15-25                   & Lower div. & L-mode density ramp  \\
45121     & 750        & 470            & 1               & 30-35                   & Midplane       & H-mode               \\
45370     & 450        & 430            & 2               & 45-70+                  & Midplane       & L-mode (high-n)
\end{tabular}
\label{tab:Overview}
\caption{Table showing main discharge parameters, including plasma current ($I_p$), power crossing the separatrix ($P_{SOL}$), up/down assymmetry (mapped to the midplane) $dr_{sep}$), core Greenwald fraction $f_{GW}$, Fuelling location and a description.}
\end{table}

\begin{figure}
\centering
\includegraphics[width=0.8\linewidth]{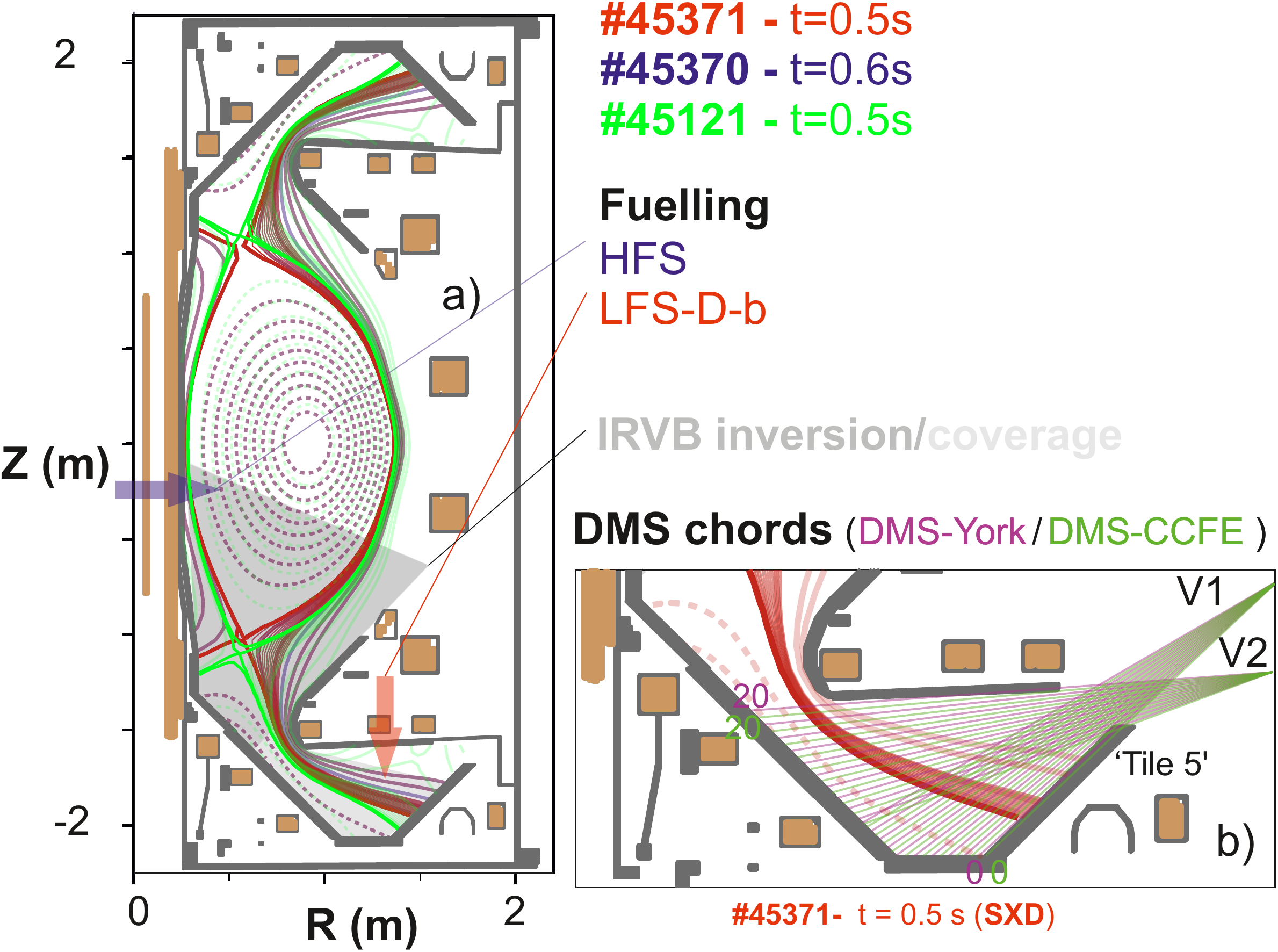}
\caption{a) The magnetic geometries corresponding to three discharges 
	\# 45371 (red) at 500 ms and \# 45370 (blue) at 500 ms and \# 45121 (green) at 500 ms, are shown together with the vessel geometry, poloidal field magnets and the fuelling valve locations utilised (`HFS' and `LFS-D-b'). The red and blue equilibria are similar, such that both overlap into dark magenta. The shaded region indicates the IRVB coverage (light grey) and where a 2D inversion of the radiation from the IRVB can be obtained (dark grey). b) Just the lower divertor region is shown along with the DMS spectroscopic chordal lines-of-sight originating from view points V1 and V2, which are both coupled to two spectrometers ('DMS-York' and 'DMS-CCFE'). In both a) and b), the Super-X separatrix strike point is incident on 'Tile 5'.}
\label{fig:GeomLoS}
\end{figure}

In this work we will mainly discuss results from Ohmic Super-X Double Null diverted plasmas in L-mode (\# 45371) in section \ref{ch:resultsOhmic} and H-mode (\# 45121) in section \ref{ch:Hmode}. The default setup of the lower Divertor Monitoring Spectrometer (DMS) \cite{Verhaegh2023} was used for both discharges, which facilitates BaSPMI analysis \cite{Verhaegh2021b} and uses two spectrometers with the following settings: 1) the $n=5,6$ Balmer lines are monitored at medium spectral resolution (0.09 nm); 2) part of the $D_2$ Fulcher band (595 - 615 nm) and $D\alpha$ (656 nm) are monitored simultaneously at low spectral resolution (0.4 nm). High-n $n\geq9$ Balmer line measurements from a deeply detached L-mode discharge (\# 45370) are used in section \ref{ch:Te_checks}. The magnetic geometries are shown in figure \ref{fig:GeomLoS}, together with the spatial coverage of the DMS and imaging bolometer (IRVB \cite{Federici2023}, which monitors the X-point region) diagnostics. The region in which the IRVB data can be inverted to obtain 2D distributions of the radiation is reduced with respect to the absolute coverage, as there has to be a sufficient number of lines of sight (corresponding to each pixel) intersecting the poloidal plane at each location (figure \ref{fig:GeomLoS}) \cite{Federici2023}. A Multi-Wavelength Imaging (MWI) \cite{Feng2017,Wijkamp2023} diagnostic, similar to MANTIS at TCV \cite{Perek2019submitted}, is used to monitor the lower divertor, which has also one channel dedicated to Coherence Imaging Spectroscopy \cite{Allcock2021,Silburn2014}.

\begin{figure}
\centering
\includegraphics[width=\linewidth]{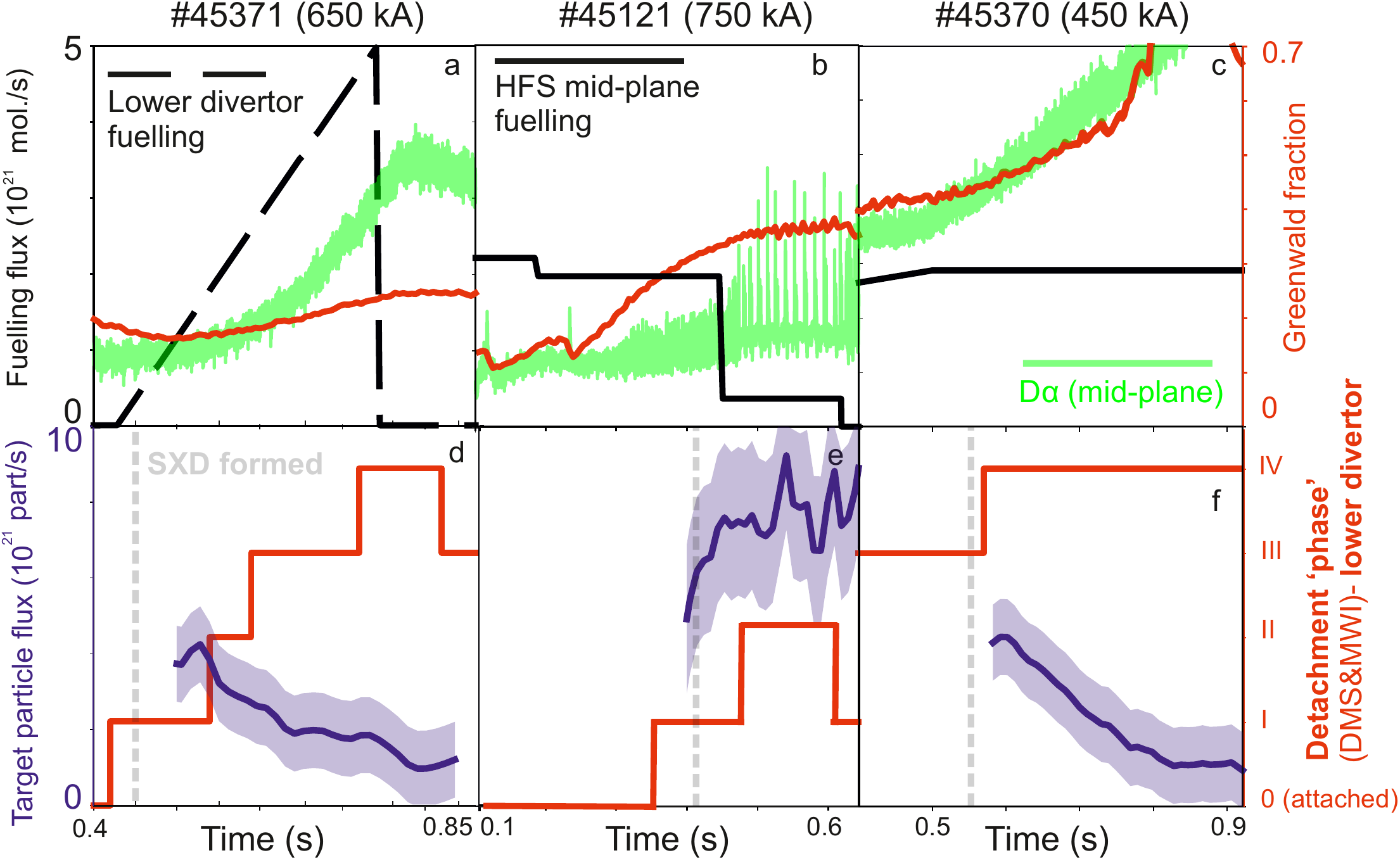}
\caption{Overview traces of discharges used in this paper of the core density (in terms of Greenwald fraction), core $D\alpha$ signal, ion target flux (lower outer target) and detachment phase (inferred spectroscopically with line-of-sight spectroscopy - DMS \cite{Verhaegh2023} and imaging - MWI \cite{Wijkamp2023}) \# 45371 (Ohmic L-mode density ramp $I_p = 650$ kA), \# 45121 (Ohmic H-mode discharge, $I_p = 750$ kA), \# 45370 (deeply detached Ohmic L-mode density ramp $I_p=450$ kA). The time at which the SXD geometry is formed is indicated by a grey dashed vertical line. The target particle flux is only shown for time ranges in which the SXD geometry was fully formed and held constant.}
\label{fig:Overview}
\end{figure}

Key parameters during the discharge, including core density, $D\alpha$ photomultiplier tube measurements, fuelling and the ion target flux \footnote{The ion target flux magnitudes have large uncertainties, since 1) only a part of the spatial profile is measured due to a lack of coverage in most cases, which is interpolated over and then integrated to estimate the ion target flux.; 2) strike point splitting in \# 45371 and \# 45121 causes toroidal asymmetries. Therefore, the indicated uncertainties apply to the relative trends of the ion target flux measurements, which are more reliable than their magnitudes ($\times \sim2-3$ uncertainty, based on up/down asymmetries in expected up/down symmetric conditions when one divertor only has partial Langmuir probe coverage) \cite{Moulton2023}.} as well as spectroscopically inferred detachment phases are shown in figure \ref{fig:Overview}. \# 45371 is fuelled from the lower divertor chamber, whereas \# 45370 \& \# 45121 are fuelled from the high-field side, main chamber. \# 45371 utilises a cut in the fuelling near the end of the discharge to monitor the divertor response to a lack of fuelling in deep detached conditions. \# 45121 enters ELM-free H-mode at $t=0.24$ s and transitions to ELMy H-mode at $t=0.46$ s as the fuelling is reduced. 

% The fuelling ramp in \# 45371 scans the divertor from near the detachment onset to very deep levels of detachment where the electron-ion recombination front (which coincides with the electron density front \cite{Verhaegh2023}) ultimately detaches from the target: e.g. it exhibits all four phases of detachment (figure \ref{fig:DetachmentPhases}). The lower current \# 45370 discharge performs a fuelling ramp starting from a deeply detached reference (similar to the deepest level of detachment obtained in \# 45371), where ultimately the electron density bulk starts to move out of the divertor chamber and a MARFE is triggered. The H-mode discharge remains detached in phase I-II. 

Measurements of $D\alpha$, $n=5$ and $n=6$ Balmer line brightnesses for pulses \# 45371 and \# 45121, using the lower divertor DMS, are the basis for inferring quantitative information about the power and particle sinks \& sources, using BaSPMI \cite{Verhaegh2021a} analysis. The $n=6$ Balmer line has been used to estimate the electron density through Stark broadening and a large uncertainty ($>3\times10^{19} m^{-3}$) has been assigned to this due to the relatively low spectral resolution (0.09 nm). The $D_2$ Fulcher band brightness has been used as a temperature constraint \cite{Verhaegh2023}. For this calculation, a fully Bayesian version of BaSPMI was used \cite{Verhaegh2023}. For more information on the BaSPMI implementation, see  \ref{ch:BaSPMI_MolCX}.

\section{Results from an L-mode Super-X ohmic fuelling ramp scan (\# 45371)}
\label{ch:resultsOhmic}

\subsection{Evolution of ion sources and sinks and particle balance in the Super-X divertor}
\label{ch:partbal}

\begin{figure}
\centering
\includegraphics[width=0.8\linewidth]{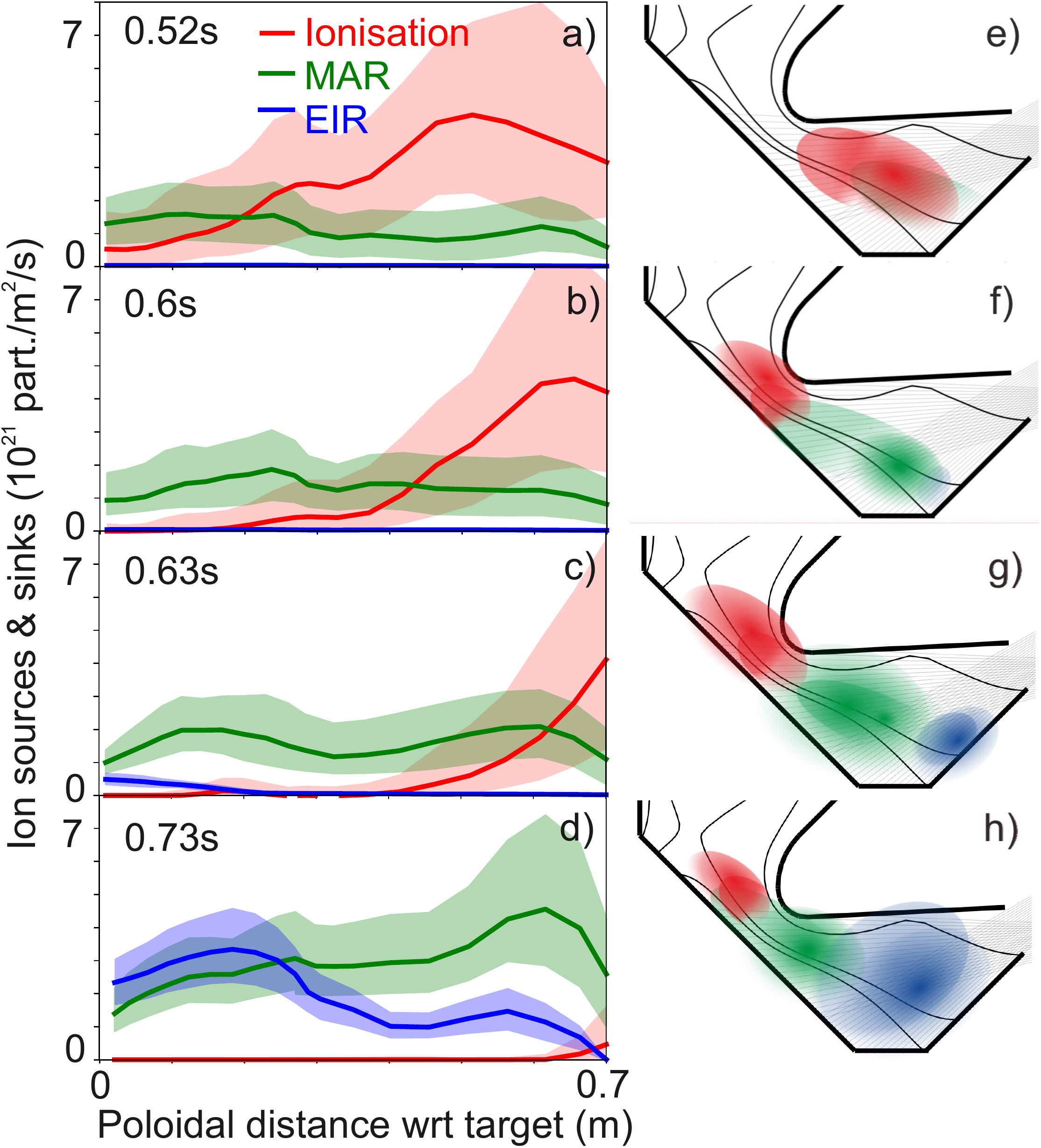}
\caption{Ion source \& sink profiles for \# 45371 (a-d) as function of poloidal distance from the target, inferred by BaSPMI, at four different time points in the Super-X divertor, indicative of the evolution of the four different phases of detachment illustrated in figure \ref{fig:DetachmentPhases}. (e-h) Schematic overview of divertor ion sources/sinks adopted from figure \ref{fig:DetachmentPhases}.}
\label{fig:PartBalProf}
\end{figure}

Figure \ref{fig:PartBalProf} displays the quantitative ion source and sink profiles inferred using BaSPMI for \# 45371 at four different time points, corresponding to detachment phases I-IV (figure \ref{fig:DetachmentPhases}). The inferred ionisation source profiles indicate that the ionisation region moves upstream as the divertor fuelling is increased and is detached from the target during the Super-X formation. According to the ion source/sink inferences (figure \ref{fig:PartBalProf}), Molecular Activated Recombination (MAR) is present throughout most of the divertor chamber and is elevated below the ionisation region. As the divertor fuelling increases, the MAR ion sink strength initially rises, ultimately moving upstream as the peak in MAR detaches from the target (detachment phase II). 
%These results confirm earlier suspicions based on the movement of $D_2$ Fulcher emission \cite{Wijkamp2023,Verhaegh2023} and the detachment of the electron-impact excitation emission from the target, which also moves upstream as the divertor fuelling is increased \cite{Verhaegh2023}.

The point where the divertor MAR ion sink exceeds the divertor ionisation source follows the movement of the ionisation region upstream. These quantitative results establish a clear spatial transition between an ionisation-dominated region and a recombination-dominated region (e.g., MAR), which commences during detachment phase I and moves upstream as detachment deepens. This cross-over point between recombining and ionising plasma is observed to be correlated with the Fulcher emission region and is expected to impact the plasma flow profile (section \ref{Implications for plasma-edge simulations}).

Detachment phase III marks the onset of electron-ion recombination, which becomes significant in detachment phase IV when the EIR peak detaches from the target. This indicates a shift in the electron density maximum away from the target, as reported in \cite{Verhaegh2023}. While the peak in the MAR ion sink moves upstream off the target in detachment phase II, significant MAR persists in the region below its peak, even in the presence of strong EIR. Temperature estimates based on EIR emission data suggest electron temperatures below or around 0.2 eV, according to several analyses including new ADAS data for $T_e<0.2$ eV (see section \ref{ch:Te_checks}). At such low temperatures, strong MAR ion sinks would not be expected according to the molecular charge exchange cross-sections used by EIRENE required for the formation of $D_2^+$ \cite{Verhaegh2023,Verhaegh2023a}, suggesting inaccuracies in these rates at low temperatures (see section \ref{ch:relevance}).

\begin{figure}
\centering
\includegraphics[width=0.7\linewidth]{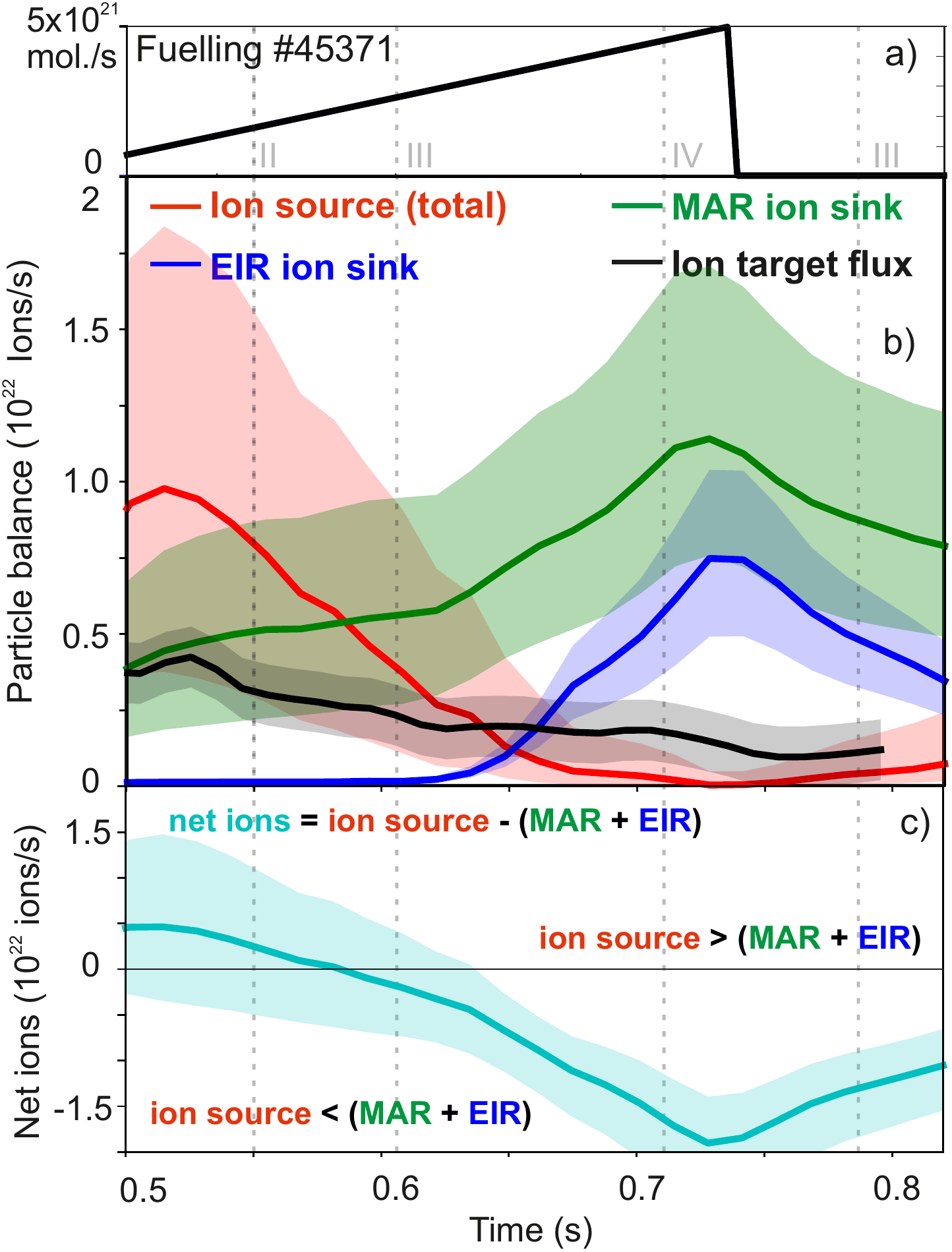}
\caption{Particle balance for \# 45371 in the Super-X divertor. a) fuelling reference trace; b) integrated ion sources \& sink profiles (figure \ref{fig:PartBalProf}) in the lower divertor compared with the ion target flux measured by Langmuir probes. c) net volumetric ion source ($>0$)/sink ($<0$) in the divertor chamber obtained spectroscopically by subtracting the total ion sink (MAR \& EIR) from the ion source. Vertical lines have been added to indicate transitions in detachment state.}
\label{fig:PartBal}
\end{figure}

Integrating the ion source/sink profiles in the divertor volumetrically, we can estimate the total ion sources/sinks (ions/s) below the baffle entrance in the lower divertor chamber (figure \ref{fig:PartBal}). As detachment progresses, the total divertor ion source reduces and the MAR ion sink increases. This coincides with the movement of the ionisation source further upstream to the divertor entrance (beyond the DMS viewing region), as depicted in Figure \ref{fig:PartBalProf}. In detachment phase II and beyond, the MAR ion sinks start to exceed the ionisation source, which is shown in figure \ref{fig:PartBal} c. Despite this, there is still a detectable amount of ion target flux reaching the outer target. This suggests a loss of 'high recycling' conditions ($I_t \approx I_i - I_r$) as we infer that $I_u$ is significant (equation \ref{eq:ParticleBalance}). This is consistent with the observation that the ionisation region moves outside of the monitored region (i.e. it moves upstream of the baffle entrance). This likely leads to an escape of neutrals to outside the monitored region (upstream of baffle entrance) that become ionised and flow back towards the target. 

Since 40 \% of the poloidal leg length from the X-point to the target is outside the monitored region (i.e. the X-point is significantly more upstream than the baffle entrance), it is uncertain whether $I_u$ arises from neutrals being ionised in between the baffle entrance and the X-point, or whether they get ionised upstream of the X-point. However, we do not observe a strong increase of the hydrogenic emission near the mid-plane in \# 45371 (in contrast to the more detached discharge \# 45370), which suggests that the neutral leakage to the mid-plane scrape-off-layer is limited. Nevertheless, additional diagnostic coverage of the baffle throat region is required to investigate the upstream ionisation further and a secondary multi-wavelength imaging system for the X-point is in development.

\subsection{Hydrogenic power losses in the Super-X divertor and molecular dissociation}

\begin{figure}
\centering
\includegraphics[width=0.7\linewidth]{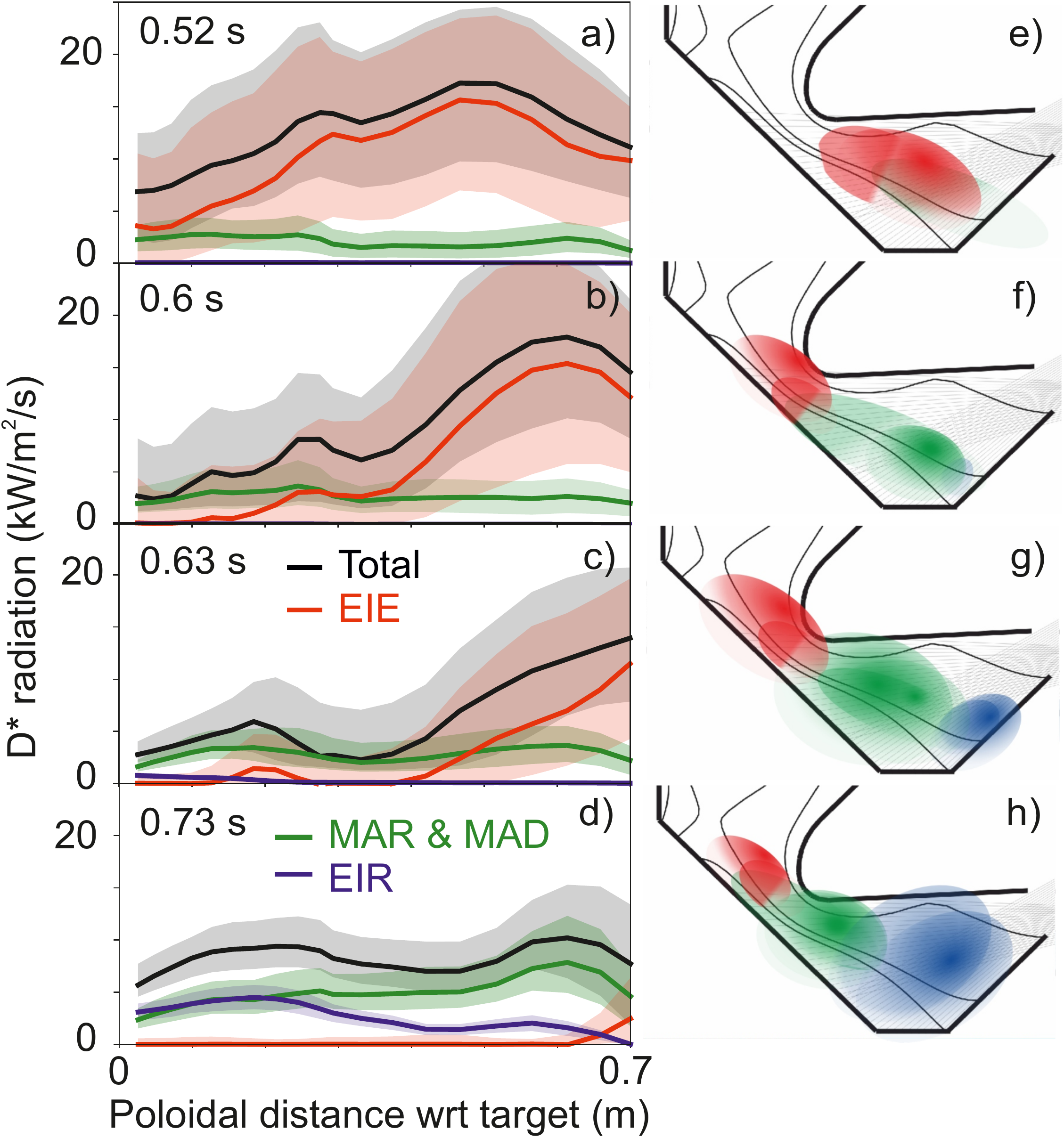}
\caption{Hydrogenic radiative loss profiles in the Super-X divertor at four different time points for \# 45371 (a-d), indicative of the evolution of the four different phases of detachment illustrated in figure \ref{fig:DetachmentPhases}. (e-h) Schematic overview of divertor ion sources/sinks adopted from figure \ref{fig:DetachmentPhases}.}
\label{fig:PradProfs}
\end{figure}

Our spectroscopic analysis also provides an estimate of the power losses due to hydrogenic processes in the monitored divertor plasma region. In figure \ref{fig:PradProfs}, we present the atomic hydrogenic radiation profiles, which include the inferred radiation losses from all excited hydrogen atoms, generated either by exciting neutral atoms or breaking down molecules. \footnote{We exclude the radiative losses from excited molecules themselves, such as the Werner and Lyman bands \cite{McLean2019,Groth2019}. Due to a lack of divertor VUV spectroscopy, this cannot be measured directly. It also cannot be inferred directly from the hydrogen Balmer line emission, but instead likely requires detailed $D_2$ Fulcher band measurements combined with collisional-radiative modelling.} Since this discharge was fuelled from the lower divertor chamber, the hydrogenic radiative losses are likely not symmetric between both divertors (e.g. it is observed that the lower divertor detaches before the upper divertor when only the lower divertor is fuelled).

Our findings, as shown in figure \ref{fig:PradProfs}, demonstrate a clear upstream movement of the electron-impact excitation (EIE) radiation region, consistent with the observed movement of the ionisation source (figure \ref{fig:PartBalProf}). Hydrogenic radiation from plasma-molecular interactions is distributed throughout the divertor. Hence, we expect the total hydrogenic radiative losses to peak near the ionisation region and move upstream with the ionisation source. As the ionisation source and hydrogenic radiative losses move out of the divertor chamber, the imaging X-point bolometry system - the IRVB \cite{Federici2023} - detects an increase in the total radiative losses above the divertor entrance. 

\begin{figure}
\centering
\includegraphics[width=0.65\linewidth]{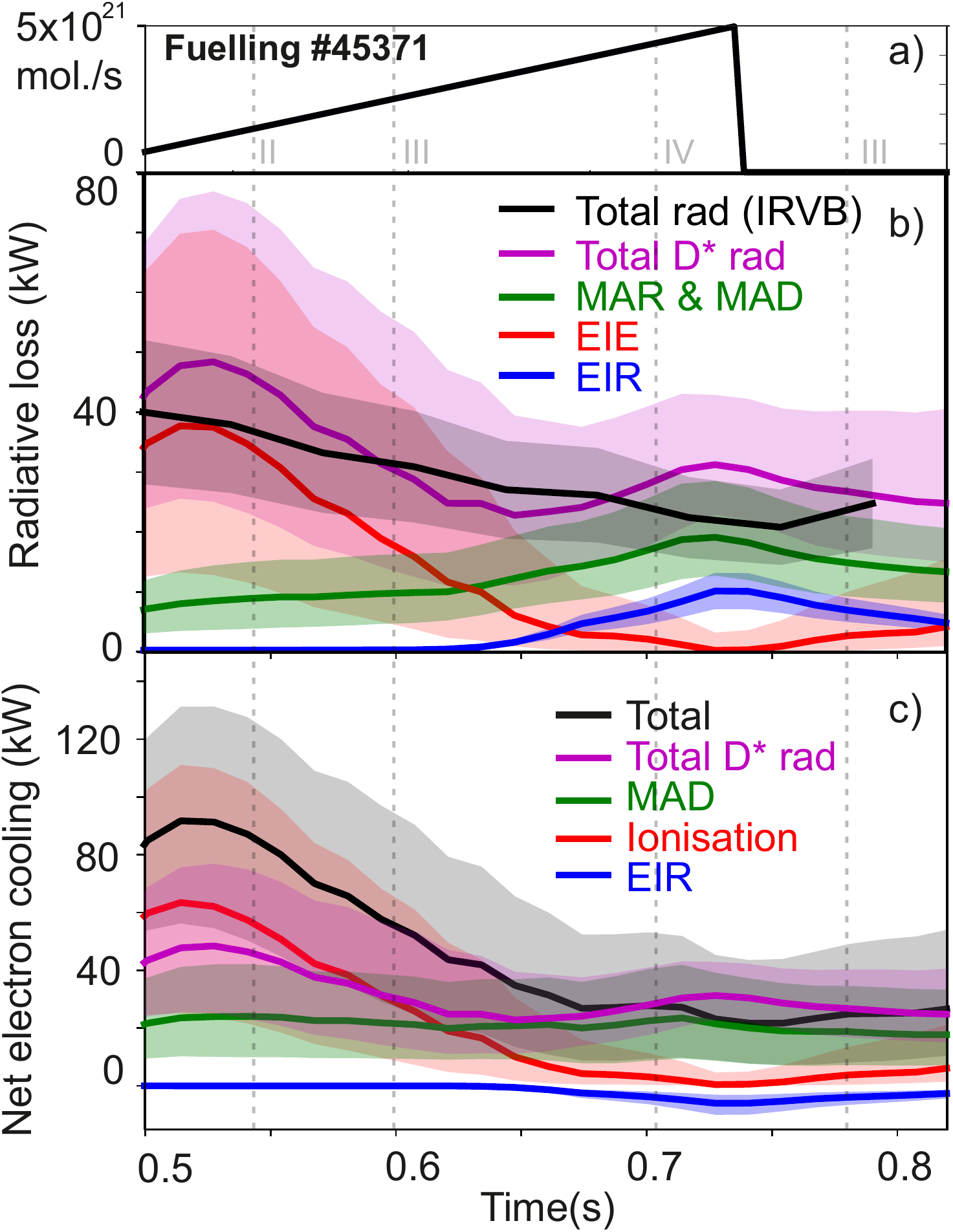}
\caption{Power losses integrated over the lower divertor chamber. a) Reference fuelling profile; b) integrated hydrogenic radiative losses based on the inferred radiative losses from BaSPMI (figure \ref{fig:PradProfs}) compared against total radiative power losses in the divertor chamber measured by the IRVB (black) c) net electron cooling power (radiative losses as well as energy gains associated with potential energy gains \& losses).}
\label{fig:Ploss}
\end{figure}

By integrating the profiles of hydrogenic radiative loss in the lower divertor chamber, we can estimate the total hydrogenic radiative losses and their various contributors (as shown in Figure \ref{fig:Ploss}). Although EIE radiative losses dominate the radiation profile, the total MAR \& MAD hydrogenic radiative losses can still be significant, forming a dominant contribution to the total hydrogenic power loss after the ionisation source moves upstream of the divertor chamber (during detachment phase II-III, $t>0.625$ s). During the deepest states of detachment (detachment phase IV), radiative power losses associated with EIR may become non-negligible. However, one may not want to operate a reactor in such a deeply detached scenario when the bulk density moves away from the target \cite{Verhaegh2023}.

Although the IRVB \cite{Federici2023} in the first MAST-U campaign could not reconstruct the 2D radiative emissivity profile in the divertor chamber, it can provide an estimate of the integrated radiation in the divertor chamber below the baffle entrance \cite{Federici2023a}. The IRVB facilitates a direct comparison against the divertor chamber integrated hydrogenic (DMS analysis) radiative losses (figure \ref{fig:Ploss}). Both results are in agreement within the substantial uncertainties. This suggests the dominant part, if not all, of the \emph{divertor chamber} radiative losses arise from hydrogenic radiative losses, which is consistent with interpretative SOLPS-ITER simulations \cite{Moulton2023}.

In addition to radiative power losses, reactions in a plasma can result in effective electron cooling or heating through transfer of potential energy. For example, ionising a neutral 'costs' $E_{ion}$ of energy to the electrons, which includes the potential energy to convert a neutral into an ion ($\epsilon$). Therefore, the electron cooling power from plasma-neutral interactions can differ from 'only' the radiative losses. However, if the ion after ionisation reaches the target, that same potential energy ($\epsilon$) is released back to the target. Hence, there is no net power dissipation from $\epsilon$ unless the ion volumetrically recombines before reaching the target. This underscores the importance of reducing the ion target flux, i.e., detachment, through ion sinks or power limitation (see section \ref{ch:introduction}).

Considering these factors, the total \emph{net} electron cooling power due to hydrogenic processes is depicted in figure \ref{fig:Ploss}c. During detachment phases I-II, the net electron cooling power is 50-100\% higher than hydrogenic radiative losses, owing to ionisation power losses. $P_{sep}$ is around 470 kW, which would imply (assuming 1:1 up/down symmetry and that no power flows towards the inner targets) that around 235 kW goes towards the lower Super-X chamber. In that case, the maximum inferred net electron cooling in the lower divertor baffled region (up to $\sim 100$ kW, figure \ref{fig:Ploss} c) can contribute up to $\sim 43$ \% of the total power going towards the lower divertor, which suggests that significant power losses occur outside the baffled region (e.g. divertor entrance and above), which increases as detachment proceeds. This is in agreement with the IRVB, as the total radiated power between x-point and the divertor entrance increases from $\sim 40$ to $\sim 80$ kW from phase I to IV detachment. 

As detachment proceeds, the power loss associated with ionisation reduces in the baffled region and electron cooling associated with MAD becomes more dominant (detachment phase II-IV) \footnote{Electron cooling associated with MAR is negligible as the hydrogenic radiative losses approximately cancel with the potential energy gained in the recombination process during MAR \cite{Verhaegh2021b}.} and can reach up to 20 \% of the power going towards the lower divertor. Despite the low electron densities ($n_e \sim 10^{19} m^{-3}$ \cite{Verhaegh2023,Clark2021,Clark2022}), our analysis shows plasma heating from EIR occurs ($\sim 15$ kW) due to the very low electron temperatures ($T_e < 0.3$ eV \cite{Verhaegh2023} - see section \ref{ch:Te_checks}).

\begin{figure}
\centering
\includegraphics[width=0.6\linewidth]{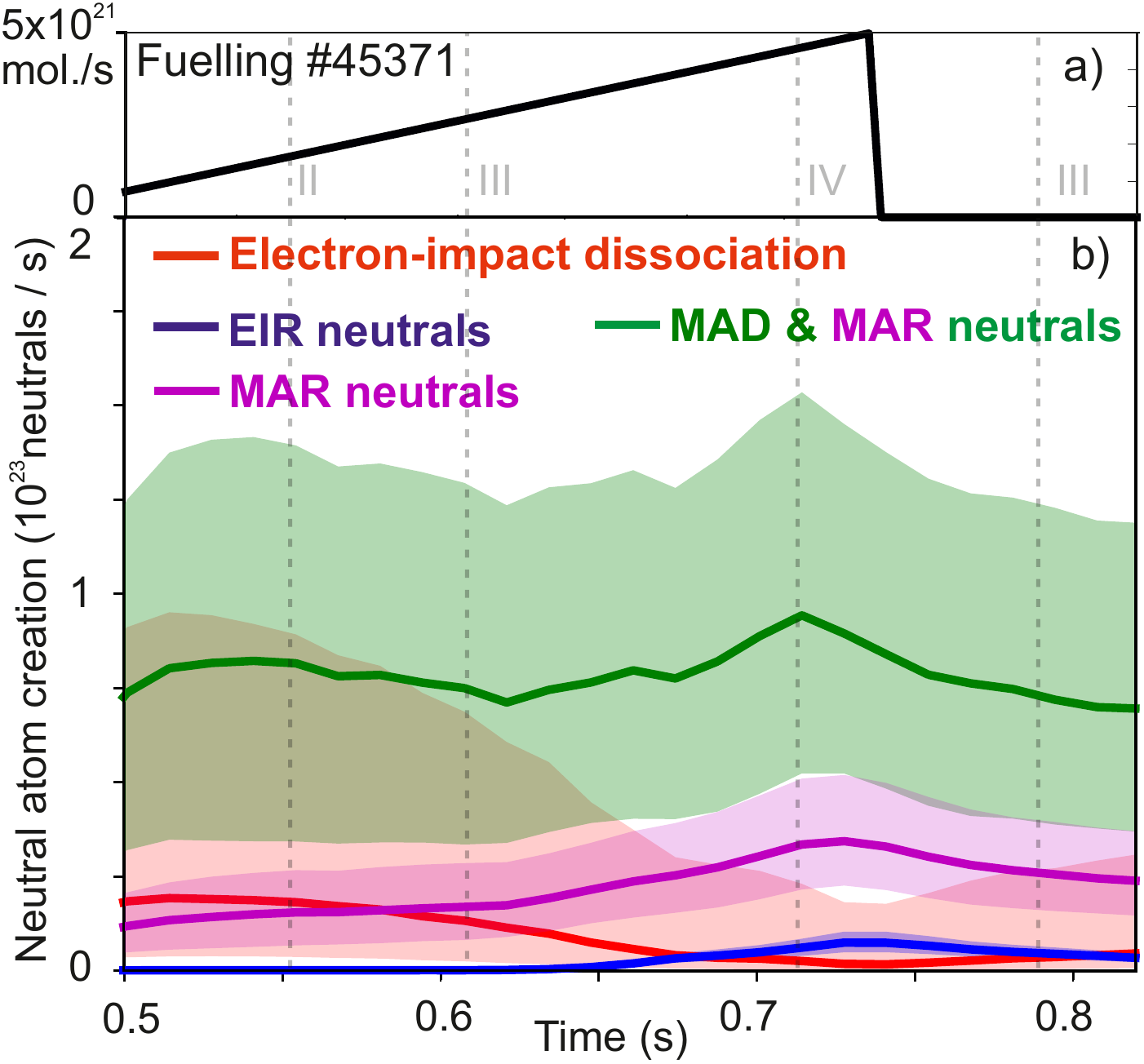}
\caption{Total volumetric neutral atom source (/s) in the lower divertor chamber inferred spectroscopically. This includes the generation of neutral atoms through electron-impact dissociation of $D_2$ \cite{Verhaegh2019a}, electron-ion recombination, Molecular Activated Recombination and the combination of MAR \& Molecular Activated Dissociation.}
\label{fig:NeutralAtoms}
\end{figure}

Plasma-molecular reactions involving $D_2^+$ and, potentially, $D^-$, not only result in ion sinks \& sources, but also lead to the generation of additional neutral atoms through MAD as well as MAR. Inferences of the volumetric creation processes of neutral atoms, integrated over the lower divertor chamber, are shown in figure \ref{fig:NeutralAtoms}. This shows that, throughout the entire discharge, MAD is the dominant neutral atom generation process. This is in agreement with previous findings on TCV \cite{Verhaegh2021a}, JET \cite{Karhunen2023} as well as SOLPS-ITER modelling with modified rates for TCV \cite{Verhaegh2023a}.

\subsection{Detachment evolution during a fuelling stop}

\begin{figure}
\centering
\includegraphics[width=\linewidth]{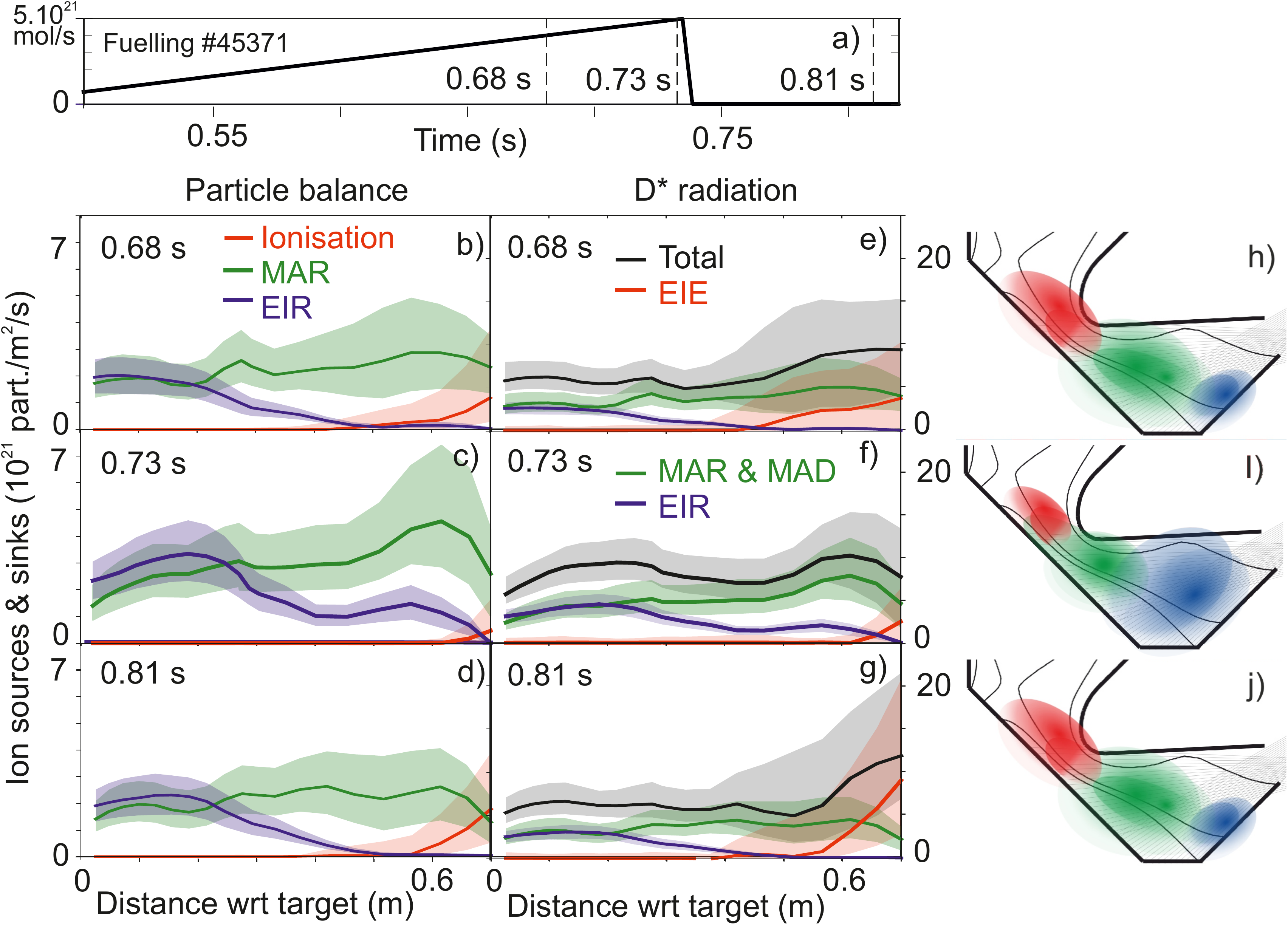}
\caption{Ion source/sink profiles (b,c,d) \& radiation profiles (e,f,g) during a gas cut (fuelling trace in a) with schematic of the detachment processes (h,i,j)}
\label{fig:GasCut}
\end{figure}

As explained in section \ref{ch:MASTU_overview}, the fuelling in the lower divertor is stopped for \# 45371 at $t\simeq 0.73$ s to monitor how a deeply detached divertor evolves during a loss of fuelling. In \cite{Verhaegh2023}, it was shown that the emission associated with EIR re-attaches at the target after the fuelling stop occurs. Spatial profiles of ion sources \& sinks as well as the hydrogenic radiation are shown in figure \ref{fig:GasCut} before, at and after the fuelling stop, with a schematic illustration of the relevant detachment processes. The evolution of the divertor chamber integrated ion sources/sinks, hydrogenic power losses and volumetric neutral atom creation mechanisms during the fuelling stop can be observed in figures \ref{fig:PartBal}, \ref{fig:Ploss}, \ref{fig:NeutralAtoms}. 

Within 80 ms from the stop of divertor fuelling: 1) the ionisation source moves back slightly towards the target (together with the total hydrogenic radiation region); 2) the peak in MAR moves downstream from the divertor entrance; 3) the EIR peak re-attaches at the target, suggesting that the electron density front re-attaches at the target and that the divertor transitions back from detachment state IV into detachment state III. This is consistent with 1) a reduction of the MAR \& EIR ion sink strength (figure \ref{fig:PartBal}); 2) an increase in the divertor ionisation source (figure \ref{fig:PartBal}); 3) a decrease in the divertor power losses associated with MAR \& MAD (figure \ref{fig:Ploss}). Although the electron density region re-attaches at the target, the divertor remains deeply detached throughout the fuelling stop phase until the end of the discharge (100 ms). A longer fuelling stop is required to test how long it would take for the divertor to re-attach.

\section{Results from a detached Super-X H-mode discharge (\# 45121)}
\label{ch:Hmode}

The discharge studied thus far (\# 45371) has been chosen as it spans a large fraction of the detached operational regime (e.g., detachment phases I to IV). However, the observations obtained are general and have been observed also in ELMy Ohmic H-mode plasmas. One caveat is that the acquisition frequency of the spectroscopy system is insufficient to capture inter-ELM periods, which means that the brightness measurements are effectively averaged over both the ELM and inter-ELM periods.

\begin{figure}
\centering
\includegraphics[width=0.8\linewidth]{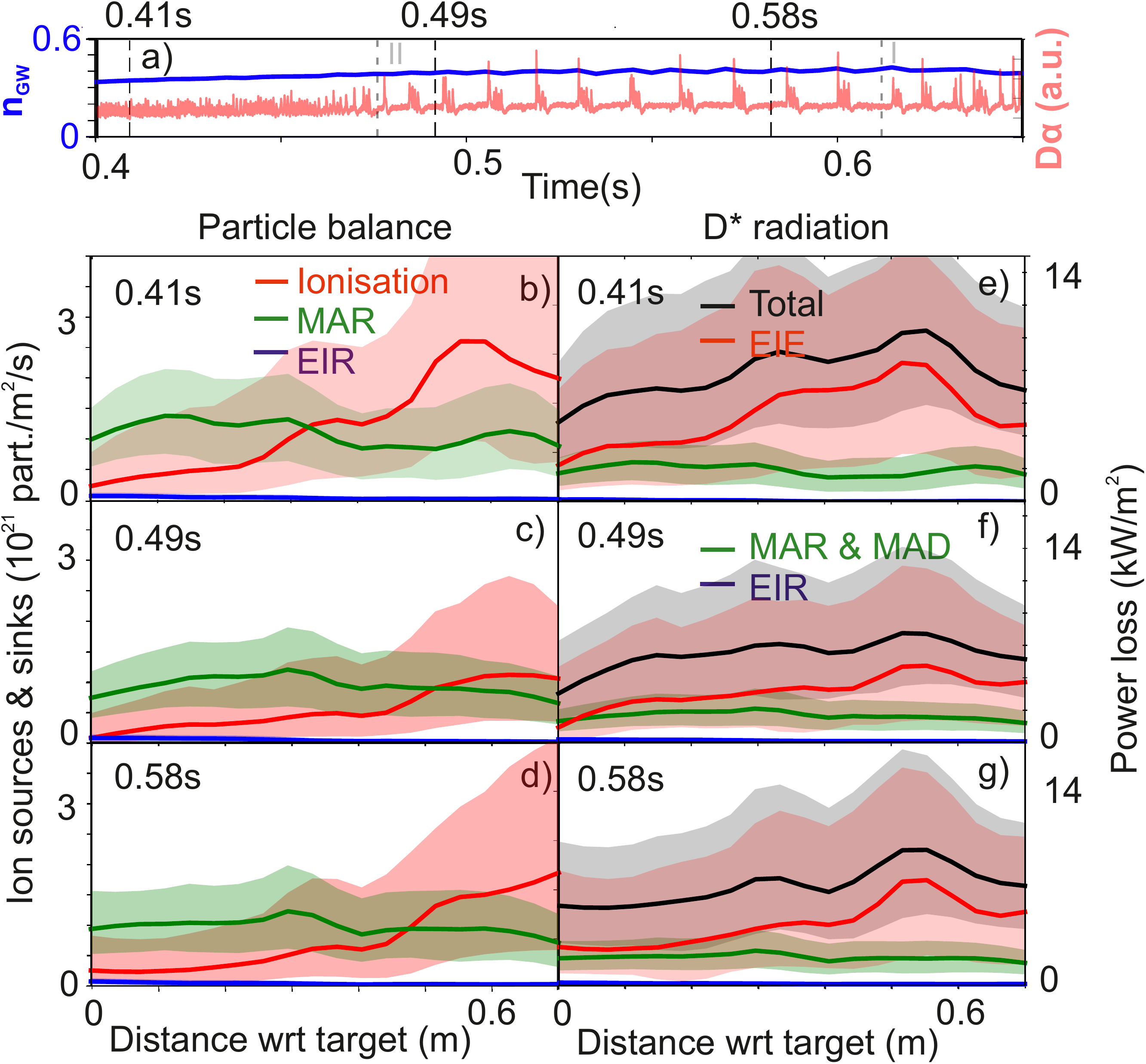}
\caption{Power and particle loss profiles for Super-X H-mode - discharge \# 45121. a) Evolution of core density (in terms of Greenwald fraction) and core tangential $D\alpha$ signal to monitor ELM activity, indicating a transitioning between an ELM-free H-mode and a type-I ELMy H-mode phase at $t=0.465 s$, with a vertical line indicating the three chosen time points. Vertical grey dotted lines are added to indicate the transition of detachment phase I to II and from II to I (adopted from figure \ref{fig:Overview}). b-d) Spatially resolved 1D ion sources/sink profiles; e-g) Spatially resolved 1D hydrogenic (e.g. $D^*$) radiative losses - both inferred from BaSPMI analysis of spectroscopic measurements. The first profile (b,e) corresponds to the ELM-free H-mode phase of the discharge. The second profile (c,f) corresponds to just after the transitioning between the ELM-free and type-I ELMy H-mode phase. The third profile (d,g) corresponds to later in the type-I ELMy H-mode phase of the discharge. }
\label{fig:HmodeSXD_PartPowerBalProf}
\end{figure} 

The spatial profiles of the divertor ion sources \& sinks, as well as the hydrogenic radiative losses, are shown at three different times in figure \ref{fig:HmodeSXD_PartPowerBalProf}. The ionisation source is strongly detached from the target for all three different profiles, with MAR being significant downstream of the ionisation source and being slightly detached from the target. No significant presence of EIR is detected in these discharges. The hydrogenic radiation spatial profile is dominated by electron-impact excitation and exhibits a similar profile to the ionisation source. This observation is, qualitatively, comparable in terms of both divertor ion sources/sinks as well as hydrogenic radiative losses to the description of detachment phase II ($\sim 0.6$ s) of \# 45371 (figures \ref{fig:PartBalProf} \& \ref{fig:PradProfs}). The inferred electron density from Stark broadening (not shown) is slightly higher for the $I_p=750$ kA H-mode discharge discussed in this section ($\sim 3.3 \times 10^{19} m^{-3}$, \# 45121) than for the $I_p=650$ kA L-mode discharge discussed in section \ref{ch:resultsOhmic} ($1.7 - 2.2 \times 10^{19} m^{-3}$, \# 45371). However, this difference is significantly smaller than the uncertainty in the Stark broadening inferences ($>2 \times 10^{19} m^{-3}$). The core density is rising in the ELM-free H-mode phase, leading to more detached conditions and a movement of the ionisation front $\sim 10$ cm further upstream, with negligible changes to the MAR ion sink profile. No significant change in the spatially resolved profiles occurs in the type-I ELMy H-mode phase, where the core density is roughly constant between $0.49$ s and $0.58$ s. 

\begin{figure}
\centering
\includegraphics[width=0.55\linewidth]{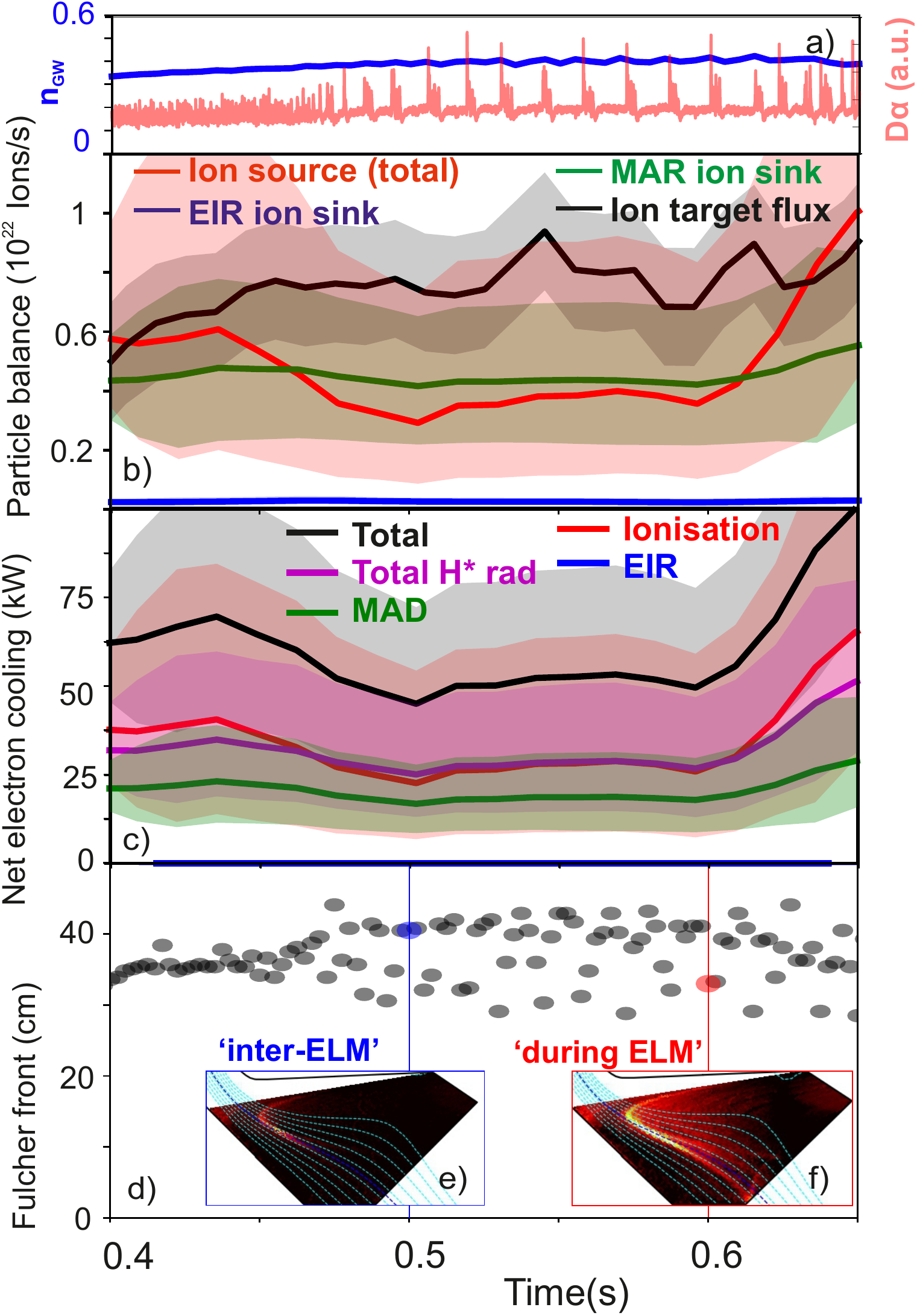}
\caption{Ion sources/sinks and hydrogenic power losses integrated over the lower divertor chamber for \# 45121. a) Evolution of core density (in terms of Greenwald fraction) and core $D\alpha$ signal to monitor ELM activity. b) Integrated ion sources/sinks using BaSPMI analysis of spectroscopic measurements. c) Integrated net hydrogenic electron cooling using BaSPMI analysis of spectroscopic measurements. d) Time trace of 50 \% peak position of the Fulcher front along the separatrix with respect to the target in the poloidal plane \cite{Wijkamp2023}, with t=0.5 s (inter-ELM, blue) and t=0.6 s (during an ELM, red) highlighted. e,f) $D_2$ Fulcher emissivity inversion at t=0.5 s and t=0.6 s respectively with indicated magnetic geometry.}
\label{fig:HmodeSXD_PartPowerBal}
\end{figure}

Integrating these spatial profiles, we obtain estimates for the total ion sources/sinks as well as hydrogenic power losses, shown in figure \ref{fig:HmodeSXD_PartPowerBal}. Ion sources and sinks, as well as hydrogenic power losses, are roughly constant in the type-I ELMy H-mode phase between 0.46 s and 0.6 s. The ionisation source is lowered when transitioning between ELM-free H-mode and type-I ELMy H-mode, which is consistent with the more detached conditions obtained at higher densities highlighted by a further movement of the ionisation upstream (figure \ref{fig:HmodeSXD_PartPowerBalProf}). No significant changes in $P_{sep}$ have been detected between the ELM-free and type-I ELMy H-mode phases. After 0.6 s, the ELM frequency seems to increase, correlated with a decrease in the core density. According to both the DMS and MWI observations, this may make the plasma slightly less detached as the ionisation source is increased and the Fulcher emission moves slightly closer to the target.

Figure \ref{fig:HmodeSXD_PartPowerBal} e,f shows an inversion of the $D_2$ Fulcher emissivity of the MWI diagnostic inter-ELM ($t=0.5$ s) and during an ELM ($t=0.6$ s) with an exposure time of 2.2 ms. The time evolution of the 50 \% front position of the Fulcher emission with respect to the target in the poloidal plane is also shown as function of time (figure \ref{fig:HmodeSXD_PartPowerBal} d). This suggests a movement of the inter-ELM Fulcher emission upstream after the transition to type-I ELMy H-mode, which is in agreement with the spectroscopic inferences in figures \ref{fig:HmodeSXD_PartPowerBalProf} and \ref{fig:HmodeSXD_PartPowerBal}. At this point, a bifurcation of the front position exists between a value that is higher up (inter-ELM) and a lower value (during an ELM). The $D_2$ Fulcher emission front, which is a proxy for the ionisation source \cite{Verhaegh2023}, does not seem to reach the target, even during an ELM. This may suggest that the ionisation source is not burning through the deeply detached divertor during an ELM in these Ohmic H-mode conditions. However, that cannot be stated with any certainty as some part of the exposure time still corresponds to the inter-ELM phase. Further investigations at higher power and with diagnostics that can perform temporal resolved measurements during an ELM are required and an ultrafast divertor spectroscopy system is in development.

\section{Discussion}
\label{ch:discussion}

\subsection{Evidence for sub-eV temperatures in the MAST-U Super-X divertor}
\label{ch:Te_checks}

\begin{figure}
\centering
\includegraphics[width=0.6\linewidth]{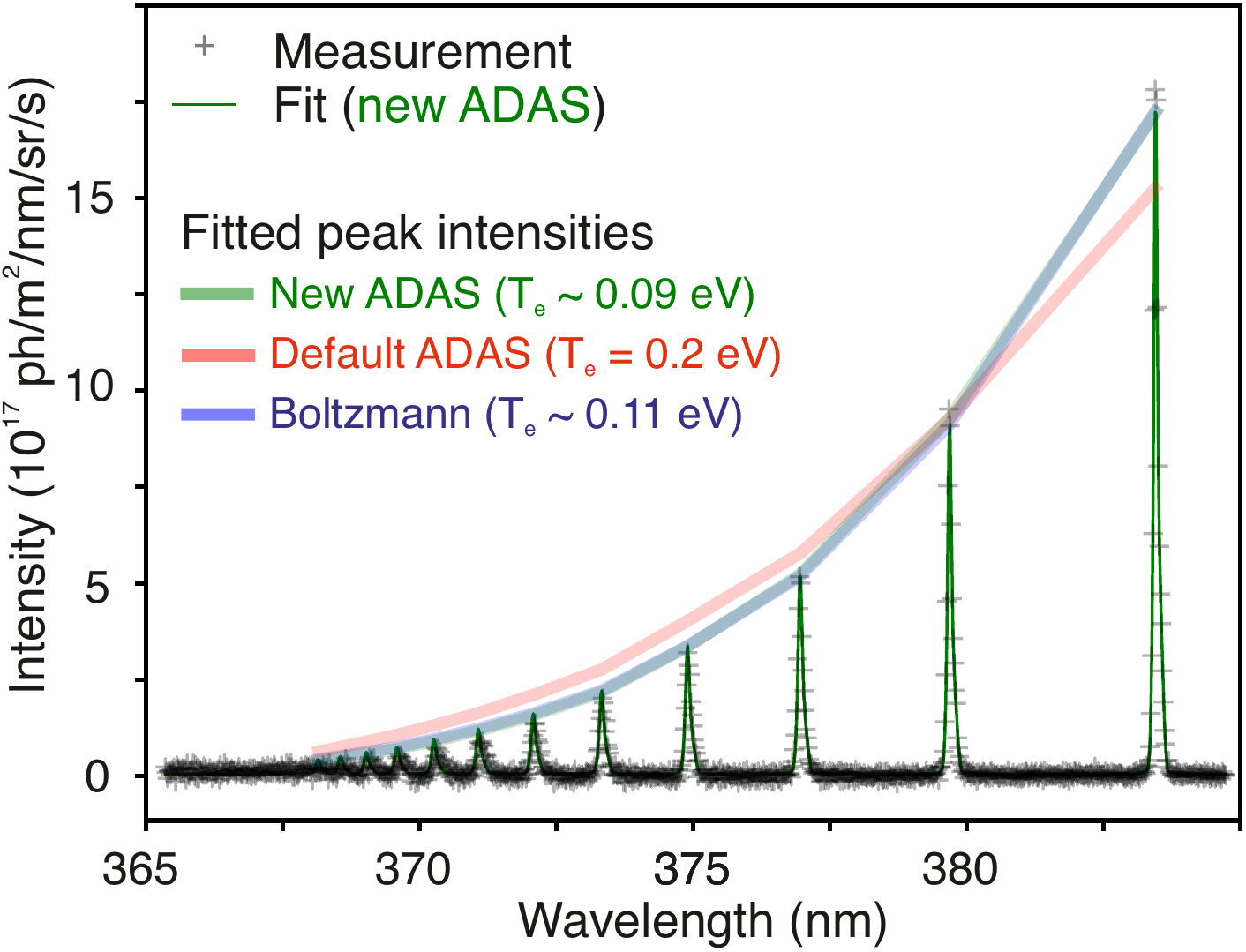}
\caption{High-n Balmer line spectra from \# 45370 from line-of-sight 6 (see figure \ref{fig:GeomLoS}) at $t=0.8$ s with example high-n Balmer line fit using new ADAS data that can go below 0.2 eV ($T_e = 0.09$ $(0.08 - 0.1)$ eV, $n_e = 3.5\times 10^{18} m^{-3}$). High-n Balmer line fits using standard ADAS data ($T_e = 0.2$ eV, $n_e = 5.1\times 10^{18} m^{-3}$) as well as a Boltzmann relation ($T_e = 0.11$ $(0.1-0.11)$ eV, $n_e = 3.7 \times 10^{18} m^{-3}$) have also been performed, but, for clarity, only peak intensities are shown in the figure rather than the full fit. The line of sight used is significantly downstream of the electron density bulk ($n_e \sim 1.4 \times 10^{19} m^{-3}$). Transparent lines are shown connecting the peaks of the various Balmer lines for the three different fits, showing an overlap between the Boltzmann fit and fit with new ADAS data and a mismatch between the measurement and the default ($T_e = 0.2$ eV) ADAS data.}
\label{fig:Highn_fit_Te}
\end{figure}

Previous research \cite{Verhaegh2023} provided evidence for $T_e$ near or below 0.2 eV in the electron-ion recombination region. However, quantitative analysis required regenerating ADAS data for electron-ion recombination that can go below 0.2 eV. Three methods were used to show $T_e < 0.2$ eV: 1) including the updated ADAS data in the ionisation sinks \& source analysis; 2) fitting the high-n Balmer line spectra using the high-n Balmer line photon emission coefficients obtained from the new ADAS data; 3) using the new ADAS data combined with Stark broadening inferred electron densities, to infer the effective emission path-length and compare this against expectations based on camera data (\ref{ch:highn_brightness}).

The characteristic temperatures of the electron-ion recombination region during deep detachment were found to be around $T_e^{EIR} = 0.17 \pm 0.05$ eV using both BaSPMI and a full Bayesian approach for the L-mode discharge discussed in section \ref{ch:resultsOhmic} (\# 45371, L-mode, Ohmic, 650 kA). No solution could be found with the default ADAS data, as the brightness of the EIR emission could not be explained. 

Using ADAS photon emission coefficients to fit the high-n ($n \geq 9$) Balmer line spectra of a different, more detached discharge (\# 45370, L-mode, Ohmic, 450 kA), shows a clear mismatch between model and observation when the default ADAS data was used and the fitting procedure would hit the 0.2 eV ADAS limit. Using the new ADAS data, temperatures below 0.2 eV are inferred, in agreement with that obtained when a Boltzmann model is used for the fit instead \cite{Verhaegh2023}. This is shown in figure \ref{fig:Highn_fit_Te} where all three fit models are compared for a deeply detached discharge (\# 45370, 450 kA Ohmic high density discharge (phase IV detachment) - see \cite{Verhaegh2023} for more information) with high-n $(n\geq9)$ coverage of the Balmer lines. The inferred electron temperatures from EIR emission are around 0.2 eV near the onset of EIR and then decay during the fuelling ramp as higher levels of EIR are achieved. Ultimately, inferred electron temperatures down to 0.08 eV are reached below the electron density bulk (see the example in figure \ref{fig:Highn_fit_Te} where 0.09 eV is reached).

\subsection{Inferences of plasma flows}
\label{ch:plasmaflows}

In this work, we observe a clear transition between an ionising and recombining region (due to MAR) in the divertor plasma (see figure \ref{fig:PartBalProf}). Such a strong transition between recombination and ionisation is expected to have a significant impact on the ion flow. We have estimated the flow profile qualitatively, for discharge \# 45371, which is an Ohmic L-mode density ramp at 650 kA discussed in section \ref{ch:resultsOhmic}, using ion target flux measurements from Langmuir probes, spectroscopically inferred ion sources and sinks and spectroscopically estimated electron densities (for details, see \ref{ch:plasmaflows_calc}).  

\begin{figure}
\centering
\includegraphics[width=0.7\linewidth]{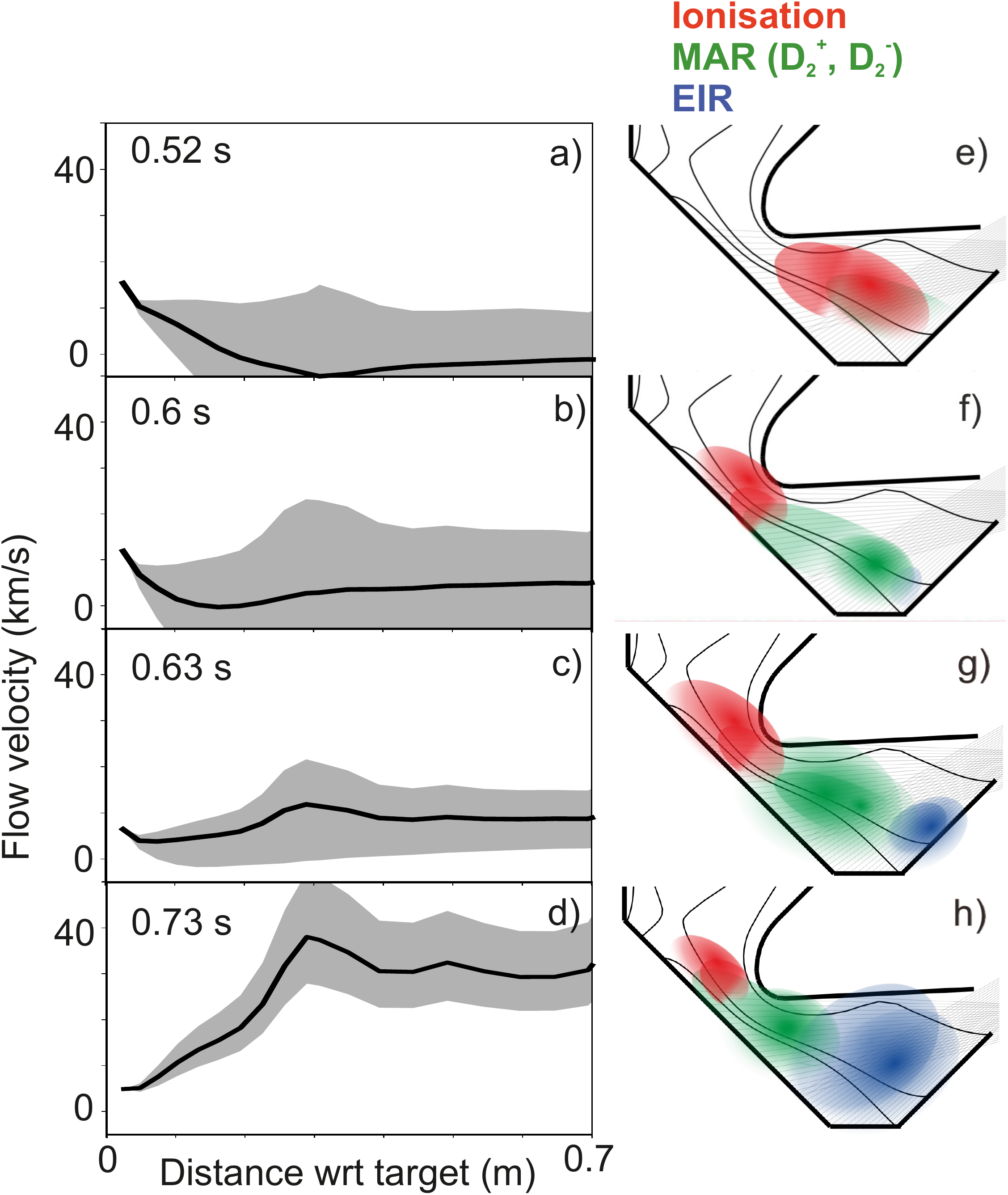}
\caption{Inferred plasma flow profile using ion sources/sinks (spectroscopy), electron density estimates (Stark broadening, spectroscopy) and ion target fluxes (Langmuir probes) for \# 45371 (a-d) at four different time points in the Super-X divertor during a fuelling scan, indicative of the evolution of the four different phases of detachment illustrated in figure \ref{fig:DetachmentPhases}. (e-h) Schematic overview of divertor ion sources/sinks adopted from figure \ref{fig:DetachmentPhases}. These calculations only include uncertainties provided by the BaSPMI analysis and, therefore, the results are only qualitative.}
\label{fig:PartFlowProf}
\end{figure}

The qualitative result obtained in figure \ref{fig:PartFlowProf} shows that the particle flow is accelerating towards the target close to the detachment onset (detachment phase I - figure \ref{fig:PartFlowProf} a). As detachment deepens, the particle flow profile flattens as ion sources and sinks start to balance in the divertor chamber (detachment phase II - figure \ref{fig:PartFlowProf} b). Ultimately, the ion flow into the divertor chamber becomes significant, increasing the flow velocity upstream, which decelerates towards the target due to the large ion sink (detachment phase III - figure \ref{fig:PartFlowProf} c, d). Such ion flow profiles are qualitatively consistent with those obtained from MAST-U SOLPS-ITER modelling \cite{Myatra2023,Myatra2023a}.

\setcounter{footnote}{0}
\subsection{Impact of fuelling location on diagnostics and detachment} 
\label{ch:Fuelling_loc}

The location of the fuelling can impact diagnostic observations as well as the physics of plasma detachment. Although there are significant differences in the spectroscopic setup of the lower and upper divertor, the onset of EIR (detachment phase III) can be compared between both divertor chambers. This occurs at similar times between the lower and upper divertor when high field side fuelling (\# 45243, \cite{Osborne2023}; \# 45370) and balanced upper/lower divertor fuelling (\# 45372) is used. However, the discussed L-mode discharge \# 45371 is fuelled from the lower divertor, in which the onset of EIR occurs first in the lower divertor. Likewise, upper divertor fuelling leads to onset of EIR first in the upper divertor (\# 45372). Therefore, our spectroscopic inferences for the lower divertor fuelled discharge \# 45371, as well as the radiative power losses, are likely up/down asymmetric. The used discharges have a slight up/down imbalance ($dr_{sep} \approx 1$ mm), which was observed (in other, high-field side fuelled, discharges) to symmetrise the upper/lower divertor particle fluxes \footnote{Upper and lower particle fluxes could not be compared for the studied discharges due to various gaps in the Langmuir probe coverage}. Therefore, we do not have any reason to assume that significant asymmetries occur in these discharges when main chamber fuelling is used.

Apart from the impact of different poloidal fuelling locations, different toroidal fuelling locations may also impact diagnostic observations as well as the physics of detachment. Additionally, as mentioned in \cite{Verhaegh2023,Wijkamp2023}, penetration of the error field as well as MHD activity caused a bifurcation of the divertor leg, leading to toroidally asymmetric strike point splitting; which was a result of the operation at low core densities (15\% Greenwald fraction).  In \cite{Wijkamp2023}, an agreement within uncertainties was found between MWI inversions and DMS measurements for midplane fuelled discharges, but not for \# 45371. The DMS lines of sight are toroidally close to the lower divertor fuelling valve. Hence, higher brightnesses were reported by the DMS as well as slightly earlier (e.g. at lower fuelling levels) appearances of the different phases of detachment, as compared to the MWI; since the DMS is impacted by the local fuelling response. Therefore, some of ion sources and sink inferences in this work may have been overestimated for \# 45371 due to toroidal asymmetries. 

Although the ion source/sink magnitudes can vary between different discharges, the reported spatial profiles of the various inferences as well as the strengths of the various inferences can be considered characteristic for these Ohmic L-mode.  Using midplane fuelled discharges only would not affect our conclusions of this work, based on BaSPMI analysis from other discharges.

\section{The implications, relevance and importance of our findings}
\label{ch:relevance}

Our results show the tightly baffled MAST-U Super-X divertor reaches unprecedented deeply detached levels.  Compared to previous TCV findings \cite{Verhaegh2021a,Verhaegh2021b} during density ramp discharges in the open divertor, the ratio between the ion target flux and the (MAR) ion sink at the deepest levels of detachment (60 \% Greenwald fraction) is similar to the least detached phase studied in this work (detachment phase I at 15 \% Greenwald fraction). The chordally integrated EIR ion sink on MAST-U at the deepest detached state, compared to TCV, is a factor four larger, despite the much lower electron density in the MAST-U divertor ($\sim 10^{19} m^{-3}$) compared to TCV ($\sim 7 \cdot 10^{19} m^{-3}$) \cite{Verhaegh2017,Verhaegh2019,Verhaegh2021}. This suggests very low electron temperatures in the MAST-U Super-X divertor, which has been supported experimentally ($T_e < 0.2$ eV). Such findings support the expectation that the MAST-U Super-X divertor leads to greatly improved divertor exhaust (i.e., power \& particle (ion) exhaust). 

That expectation is supported by the finding that hydrogenic radiation is the dominant radiative power loss \emph{in the divertor chamber}, despite the MAST-U carbon walls, which is in contrast with TCV findings \cite{Verhaegh2019}. Additional support for the improved divertor exhaust performance is provided by: 
\begin{enumerate}
\item The observation that such deeply detached conditions also exist in Ohmic H-mode operation, where the SOL width is presumably more narrow \cite{Thornton2014}, in accordance with MWI observations.
\item Deeply detached conditions do not lead to a cascading effect where the various detachment regions move further and further upstream when the (over)fuelling is stopped.
\item A loss of fuelling leads to a gradual movement of the various detachment fronts back towards the target; a loss of fuelling for 100 ms does not result in re-attachment. \footnote{Likely, detachment can be maintained without fuelling for significantly longer than 100 ms, but dedicated discharges are required to test this.}
\end{enumerate} 

Although the Super-X divertor shows strongly enhanced exhaust capabilities, it is unknown at this stage how much tight baffling and divertor closure relatively contribute to the observed divertor performance of the MAST-U Super-X and more analysis and comparison of different magnetic geometries as well as potentially cryopumping in the divertor to remove the baffle chamber neutrals is required to answer that question. Such a question should be addressed through a combination of experiments and model comparisons. 

It is also unknown how MAST-U results scale to reactor-like conditions. Validation of plasma-edge simulations for alternative divertor configurations is required on MAST-U to reduce the uncertainties in extrapolating this current knowledge to reactor-class devices. Such validation exercises are particularly challenging, however, by the deeply detached conditions of the MAST-U Super-X divertor in which processes such as MAR play a role that are not properly covered in plasma-edge simulations \cite{Verhaegh2023a}. Additionally, the Super-X divertor is particularly difficult to attach, which requires operation at low densities where strike point splitting occurs, complicating such validation exercises. Modifications in the setup of plasma-edge simulations may be required to improve their capabilities at simulating such deeply detached plasmas. 

It is currently unclear whether 1) such deeply detached conditions; 2) high MAR ion sinks are relevant for reactor designs. Whether MAR can be reactor relevant is extensively discussed in \cite{Verhaegh2023,Verhaegh2023a,Verhaegh2021a,Verhaegh2021b}. The conclusion of these discussions was that both deeply detached conditions as well as high levels are MAR are more likely in reactor designs that employ alternative divertor concepts, tightly baffled divertors and X-point radiators, which are designed to operate with the ionisation region significantly detached from the target.  Additionally, transient loss of heating in reactors could temporarily cause deeply detached conditions. Whether such interactions can be important for reactors cannot be currently answered and requires further investigations. However, even if one were to argue that such deeply detached plasmas may not, necessarily, be reactor relevant, it may be a requirement for validating the capability of simulating tightly baffled novel divertor concepts. This is required to reduce the uncertainties in extrapolating this current knowledge to reactor-class devices.

Our results indicate that the MAST-U Super-X divertor during Ohmic L-mode operation is underpowered: midway through the detachment operational regime ion sinks dominate over ion sources in the divertor and the ion target flux is driven by ion flows from upstream. Operation at higher power conditions is required to fully test the performance of the Super-X divertor.

\subsection{Implications for plasma-edge simulations}\label{Implications for plasma-edge simulations}

New ADAS data confirms previous analyses of $T_e < 0.2$ eV conditions during detachment in the MAST-U Super-X divertor. This suggests the need for re-assessment of atomic and molecular data for studies on alternative divertor configurations, both for diagnostic inferences and plasma-edge simulations, to ensure the validity of the data used in the relevant temperature regime.

Our results demonstrate that plasma-molecular interactions play a crucial role in the Super-X divertor physics of MAST Upgrade, especially in terms of additional ion sinks and neutral atom sources after detachment onset. Electron cooling power arising from MAD can also be significant (20 \% of $P_{sep}$). Interpretive SOLPS-ITER modelling underestimates the loss of ion target flux post-roll-over for MAST-U due to a lack of MAR ion sinks compared to the experiment \cite{Moulton2023}. Such observations are consistent with SOLPS-ITER modelling for TCV \cite{Wensing2019,Wensing2020,Fil2017,Verhaegh2023a}, which also lacks ion target flux roll-over and MAR ion sinks. These discrepancies arise from the underestimation of $D_2^+$ content in detached SOLPS-ITER simulations due to inaccuracies in the rates used for molecular charge exchange by Eirene \cite{Verhaegh2021b,Verhaegh2023a,Perek2022}. This work reveals the need for addressing gaps in plasma-edge modelling to reduce uncertainties when extrapolating current knowledge to reactor-class devices.

\subsubsection{The occurrence of MAR in deeply detached regimes ($T_e < 1$ eV)}

Our findings of MAR ion sinks, even in EIR-dominant plasma ($T_e < 0.2$ eV) \, disagree with EIRENE's molecular charge exchange rates \cite{Verhaegh2023,Verhaegh2023a}. Molecular charge exchange cross-section calculations that are fully vibrationally resolved indicate an almost energy independent cross-section at high vibrational levels ($\nu \geq 4$ for hydrogen) \cite{Ichihara2000,Reiter2018}. Therefore, MAR in a $T_e<0.2$ eV regime requires a sufficient fraction of highly vibrationally excited molecules reaching the cold detached region. 

\begin{figure}
\centering
\includegraphics[width=0.7\linewidth]{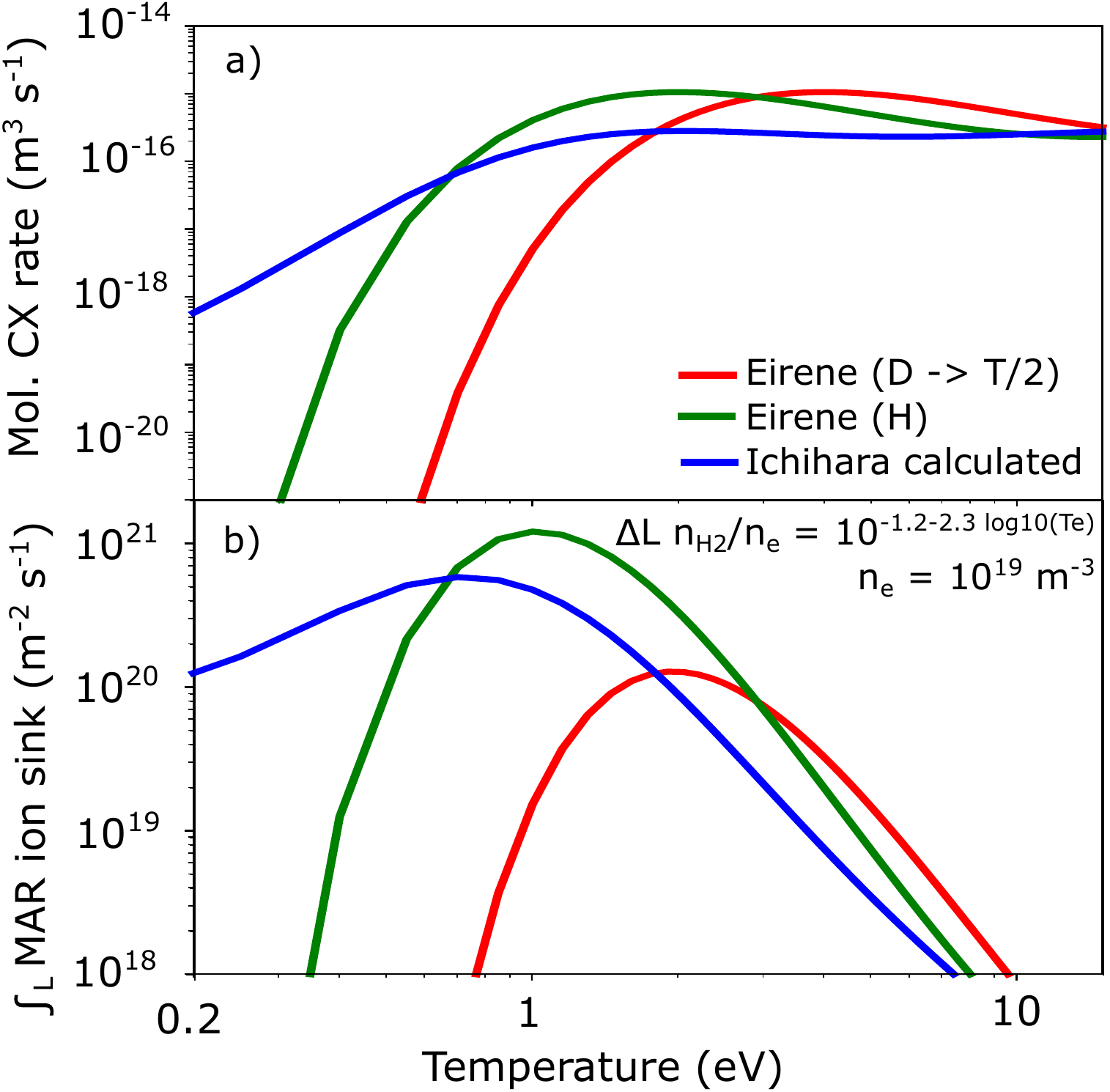}
\caption{Comparison of a) calculated molecular charge exchange and b) line-integrated MAR rates, assuming a $\Delta L n_{H_2} / n_e = 10^{-1.176 - 2.3075*log10(T_e)}$ scaling obtained from \cite{Verhaegh2023}, as function of temperature ($T_e = T_i$ assumed)}
\label{fig:MolCX_rate}
\end{figure}

Although vibrational excitation by electron impact may be less efficient at $T_e < 1$ eV \cite{Laporta2021}, collisional-radiative modelling predicts that one may expect a sufficient fraction of highly vibrationally excited molecules to sustain MAR at such temperatures. This is shown in figure \ref{fig:MolCX_rate}, where CRUMPET \cite{Holm2022} was used to model the vibrational distribution. However, the vibrationally resolved molecular charge exchange rates in the 'default' Eirene reaction set are inaccurate at low temperatures ($T<1.5$ eV for hydrogen, $T<3$ eV for deuterium) as an analytic rescaling of the rate for $\nu=0$ is employed \cite{Verhaegh2023a,Greenland2001,Janev1987,Reiter2005}. These rates have been replaced with those obtained from Ichihara, which are computed using a fully vibrationally resolved quantum mechanical calculation \cite{Ichihara2000}\footnote{See appendix \ref{ch:colrad} for more information on the collisional-radiative model implementation and see \cite{Verhaegh2023a} for more discussion on the molecular charge exchange rate employed in Eirene.}. 

The calculated molecular charge exchange rate (blue, figure \ref{fig:MolCX_rate} a) near 1 eV is greatly increased compared to the ion mass rescaled Eirene rate for deuterium. The calculated rate is of similar (although a bit lower) magnitude as the default hydrogen rate used by Eirene, in agreement with predictions in \cite{Verhaegh2023a}. There is a strong enhancement for the molecular charge exchange rate when the Ichihara rates are used at below 1 eV. Although the effective molecular charge exchange rate is reduced at low temperatures (figure \ref{fig:MolCX_rate} a), the MAR ion sink can still be significant as the molecular density increases at low temperatures. To account for this, the DMS line-of-sight integrated MAR ion sink has been modelled using the simplified rate modelling and molecular density scalings from  \cite{Verhaegh2023,Verhaegh2023a}. This shows that significant MAR rates can be sustained at low temperature conditions (figure \ref{fig:MolCX_rate} b) - in agreement with our MAST-U observations. 

Nevertheless, a significant reduction (factor 4) of MAR at low temperatures (from 0.7 to 0.2 eV, figure \ref{fig:MolCX_rate} b) is expected. However, there are additional processes that can increase the fraction of highly vibrationally excited molecules at low temperatures. First, the Eirene rates for vibrational excitation through electron-impact collisions are obtained by analytically rescaling rates for the ground states \cite{Greenland2001,Reiter2005}. This approach may underestimate vibrational excitation at low electron temperatures as it does not account for the reduced energy threshold for vibrational excitation at higher vibrational levels \cite{Laporta2021}. Secondly, vibrationally excited molecules may be generated in one part of the plasma and transported into the cold region \cite{Wischmeier2005}. Thirdly, re-distribution of vibrationally excited levels through electronically excited states \cite{Chandra2023,Holm2022} may play a role in generating vibrationally excited molecules in the $D_2$ Fulcher emission region ($T_e > 4$ eV). Fourthly, plasma-surface interactions may result in higher vibrationally excited molecules in the cold detached region. Further research is necessary regarding the vibrational distribution, using both experimental measurements ($D_2$ Fulcher band spectroscopy \cite{Osborne2023}) and modelling. Underestimates of MAR may lead to incomplete estimates of particle balance in SOLPS-ITER and affect the spatial profile of ion sources and sinks, which could lead to discrepancies in the plasma flow profile (section \ref{ch:plasmaflows}) of plasma-edge simulations. 

\subsection{Implications of the importance of hydrogenic power losses}

Our findings showed that the hydrogenic radiative power losses in the Super-X chamber are in agreement with the total measured radiative power losses by bolometry, suggesting insignificant contributions of intrinsic carbon radiative losses \emph{in the divertor chamber} despite the carbon walls on MAST-U. In contrast, TCV findings \cite{Verhaegh2019,Verhaegh2021b} show that the hydrogenic radiative losses were only $\sim 20\%$ of the total measured radiative loss by bolometry in experiments without extrinsic seeding. The difference between hydrogenic and total radiative losses was attributed to intrinsic carbon radiative losses. This difference between MAST-U and TCV is an important result that is in agreement with both TCV \cite{Wensing2019,Wensing2020,Fil2017} as well as MAST-U \cite{Moulton2023} interpretive SOLPS-ITER simulations \footnote{A 3.5 \% \cite{Wensing2019,Wensing2020} and 5 \% \cite{Fil2017} chemical erosion yield is assumed for TCV, whilst the Haasz-Davis model is used for MAST-U.}; which needs to be accounted for when comparing MAST-U and TCV results. For MAST-U, the magnitude of the CII (426 nm) brightness (DMS) and CIII (465 nm) emissivity (MWI) are in agreement within 50 \% of the simulations, suggesting that the carbon content in the simulation is similar to that of the experiment.

The finding that hydrogenic power losses are dominant in MAST-U may be a result of the tightly baffled Super-X divertor. The tight baffling concentrates plasma-neutral interactions in the MAST-U divertor chamber, amplifying hydrogenic power losses. Simultaneously, the Super-X divertor results in reduced ion target fluxes as well as a reduced detachment onset threshold, greatly reducing the ion target flux and hence - at a fixed chemical sputtering percentage - the chemical erosion of carbon would be reduced. Additionally, the electron temperature could be significantly lower in the cold MAST-U Super-X divertor conditions, which can displace or remove the carbon radiation from the divertor chambers. MAST-U operation at higher power may lead to more significant ion fluxes towards the target and more chemical sputtering, and thus a higher carbon impurity concentration and more carbon radiative losses. 

Carbon radiation plays a more significant, if not dominant, role upstream of the divertor chamber near the X-point region and the scrape-off-layer. SPRED core VUV spectroscopy indicates carbon is a dominant radiator in the main chamber for \# 45371. In the main chamber, the carbon concentration may be elevated due to main chamber erosion \cite{Wischmeier2005}. The imaging bolometer measures up to 50 kW radiation (\autoref{fig:Ploss}) in the divertor chamber at 0.5 s. Above the divertor entrance (at the inner target, SOL, X-point and divertor leg) it measures 75 kW at 0.5 s rising to 175 kW at 0.7 s. A significant portion of this radiation likely occurs electron-impact collisions preceding ionisation above the divertor entrance (\autoref{fig:PartBal}), estimated to be up to 50-100 kW during deep detachment (assuming 20-25 eV of radiation per ionisation event). Further research into the carbon concentrations and radiation upstream and in the divertor chamber is required for understanding the MAST-U power balance.

\section{Conclusions}
\label{ch:conclusion} 

Quantitative investigation of plasma-atom and molecular interactions have led to inferences of the the divertor ion sources and sinks and hydrogenic power losses. This has shown that Molecular Activated Recombination (MAR) plays an unprecedented strong role in the physics of the MAST Upgrade Super-X divertor. Our results indicate that the detached operational window of the MAST-U Super-X divertor, in terms of core density, is very large and ion sinks dominate over ion sources in the divertor chamber overall for a significant part of that operational window. This implies that ion flows from outside the divertor chamber are significant, leading to an amplified ion flow at the entrance of the divertor baffle that is decelerated as the plasma moves to the target. After the detachment onset, the ionisation source detaches from the target and MAR builds up in the entire divertor chamber. The onset of MAR occurs before electron-ion recombination (EIR) and MAR remains a dominant ion sink even when EIR becomes significant. When EIR becomes significant, electron temperature estimates below 0.2 eV are reported based on new ADAS data in combination with imaging measurements of the $n=9$ Balmer line. Spectroscopic inferences of the total hydrogenic radiative loss in the divertor are in agreement with imaging bolometry, implying that hydrogenic radiation is the dominant radiative loss mechanism, despite MAST-U featuring a carbon wall. The radiation profile is peaked near the ionisation source, where it is dominated by electron-impact excitation. However, the total power losses can have significant components from plasma-molecular interactions, which are more spread out in the divertor and lead to MAR and Molecular Activated Dissociation (MAD), which is the dominant volumetric neutral atom generation mechanism. These findings are general and are observed during Ohmic L-mode as well as Ohmic H-mode plasmas (ELM integrated). ELM temporally resolved measurements using multi-wavelength imaging indicates that the ionisation source is significantly detached from the target inter-ELM.

\section{Acknowledgements}

Discussions with Juuso Karhunen have been very helpful and have been kindly acknowledged. The results are obtained with the help of the EIRENE package (see www.eirene.de) including the related code, data and tools \cite{Reiter2005}. This work has received support from EPSRC Grants EP/T012250/1 and EP/N023846/1. This work has been carried out within the framework of the EUROfusion Consortium, partially funded by the European Union via the Euratom Research and Training Programme (Grant Agreement No 101052200 — EUROfusion). The Swiss contribution to this work has been funded by the Swiss State Secretariat for Education, Research and Innovation (SERI). Views and opinions expressed are however those of the author(s) only and do not necessarily reflect those of the European Union, the European Commission or SERI. Neither the European Union nor the European Commission nor SERI can be held responsible for them. To obtain further information on the data and models underlying this paper please contact publicationsmanager@ukaea.uk.

\section{References}
\bibliographystyle{iopart-num}
\bibliography{all_bib.bib}

\appendix

\section{BaSPMI implementation and determining MAR/MAD/MAI estimates}
\label{ch:BaSPMI_MolCX}

This work uses a fully Bayesian version of BaSPMI \cite{Verhaegh2023}. This provided similar results to the older method \cite{Verhaegh2021a}, however the full Bayesian version was more stable in detachment onset conditions. As no $D\beta$ information is available for these discharges, we cannot analyse the relative roles of $D_2^- \rightarrow D^- + D$ and $D_2^+$ and the analysis assumes that all hydrogen emission from plasma-molecular interactions arises from interactions with $D_2^+$. As explained in \cite{Verhaegh2021b}, this is not expected to have an impact on the estimates of the various ion sources \& sinks. Additional analysis using $D\beta$ measurements would be required to make any statement on the relative roles of $D_2^+$ and $D_2^- \rightarrow D^- + D$ experimentally.

The BaSPMI analysis uses visible hydrogen emission from excited atoms to infer which process led to that excited atoms. Therefore, BaSPMI is able to sense photons arising from excited atoms after $D_2^+$ interacts with the plasma, but it cannot infer which process created $D_2^+$. There are two processes that can create $D_2^+$: 1) molecular charge exchange, particularly important at lower temperatures ($D_2 + D^+ \rightarrow D_2^+ + D$) and $D_2$ ionisation ($e^- + D_2 \rightarrow D_2^+ + 2 e^-$), relevant at higher temperatures ($T_e>4$ eV). Whether an interaction with $D_2^+$ leads to MAR, MAD or MAI depends on the process that created $D_2^+$. As such, the fraction of $D_2^+$ created by molecular charge exchange must be modelled to infer the MAR, MAD and MAI magnitudes \cite{Verhaegh2019a}. Assuming there is no transport of $D_2^+$ (which is plausible, given the high reactivity and thus short lifetimes of $D_2^+$), this can be modelled using the ratio between the molecular charge exchange rate and the sum of the $D_2$ ionisation and molecular charge exchange rates. Both these two rates depend on the vibrational distribution. 

A model for the vibrational distribution has been implicitly assumed when the polynomial fit coefficients used within Eirene (AMJUEL) were derived. Inaccuracies in both the model for the vibrational distribution as well as the vibrationally resolved reaction cross-sections \cite{Verhaegh2023a} can affect the fraction of $D_2^+$ created by molecular charge exchange. Therefore, the fraction of $D_2^+$ created by molecular charge exchange is modelled using Monte Carlo uncertainty propagation. In this, vibrationally resolved molecular charge exchange rates from Ichihara \cite{Ichihara2000} are used for molecular charge exchange, which are based on vibrationally resolved ab initio quantum mechanical simulations, in contrast to the oversimplified analytically rescaled rates used by Eirene \cite{Verhaegh2023a,Holliday1971,Greenland2001,Janev1987}. Since the vibrational distribution is unknown, uncertainty propagation is used to sample random vibrational distributions with log-uniform priors on the fraction of each vibrational state. For each Monte Carlo sample, the vibrational distribution is normalised. Since the molecular charge exchange vibrationally resolved rates depend on the ion temperature, it is assumed that the ion temperature is between 80\% and 150\% of the electron temperature. The differences in the relative velocity for between $H_2$ (which is fixed according to $E_{H_2} = 0.1$ eV) and $H^+$ at different isotope masses is accounted for. Uncertainties of 200 \% on the magnitudes of the reaction rates is assumed. For $H_2$ ionisation, the vibrationally resolved rates from Eirene ('H2VIBR'), which contains polynomial fit coefficients to express those rates, are used; which is originally from \cite{Holliday1971,Greenland2001,Janev1987}. 

The result of uncertainty propagation indicates that, although there are large uncertainties on virtual all input parameters, this only causes small/negligible uncertainties in the ratio between the molecular charge exchange rate and the total $D_2^+$ creation rate, which implies negligible uncertainties in the determination of MAR / MAD and MAI when plasma-molecular interactions contribute significantly to the hydrogenic emission. This is because the temperature ($T < 2$ eV), in the regime where significant MAR / MAD / MAI occurs, is sufficiently low such that the fraction of $D_2^+$ generated through molecular charge exchange is nearly 1, despite the various uncertainties. This is in contrast with recent JET analysis that finds that MAR and MAI can cancel each other, even at very low temperatures \cite{Karhunen2023}. That difference is caused by the usage of the 'H2VIBR' rates for molecular charge exchange in \cite{Karhunen2023}, which are based on applying an analytic scaling \cite{Greenland2001} to the cross-sections for molecular charge exchange in the vibrational ground state \cite{Janev1987,Holliday1971}, resulting in potentially severely underestimated molecular charge exchange rates \cite{Verhaegh2023}. This may have resulted in underestimated MAR estimates and overestimated MAI estimates for JET in \cite{Karhunen2023} at $T<2$ eV; at such temperatures one would not expect $H_2$ ionisation to be significant \cite{Reiter2018}. 

MAI could, however, play a role outside of the region where strong emission from excited atoms after plasma-molecular interactions occurs. In that region, our analysis cannot distinguish between ionisation and MAI, given the similarity in the emission signatures between these two \cite{Verhaegh2021b}. However, SOLPS-ITER simulations for TCV both with the default rate setup and a modified rate setup where ion isotope mass rescaling of the molecular charge exchange rate was disabled, find only small ($<15 \%$) MAI contributions to the total ion source for both cases \cite{Verhaegh2023a}. Additionally, the contribution of MAI would increase if the cross-sections for $H_2$ ionisation are higher than the (H2VIBR) rates we used. Additional $H_2$ ionisation may occur through multistep processes of electrons ionising electronically excited levels \cite{Scarlett2021}, however the likelihood of electronic excitation is reduced at low temperatures.

\section{Additional evidence for $T_e<0.2$ eV based on the high-n Balmer line brightness}
\label{ch:highn_brightness}

\begin{figure}[h!]
\centering
\includegraphics[width=\linewidth]{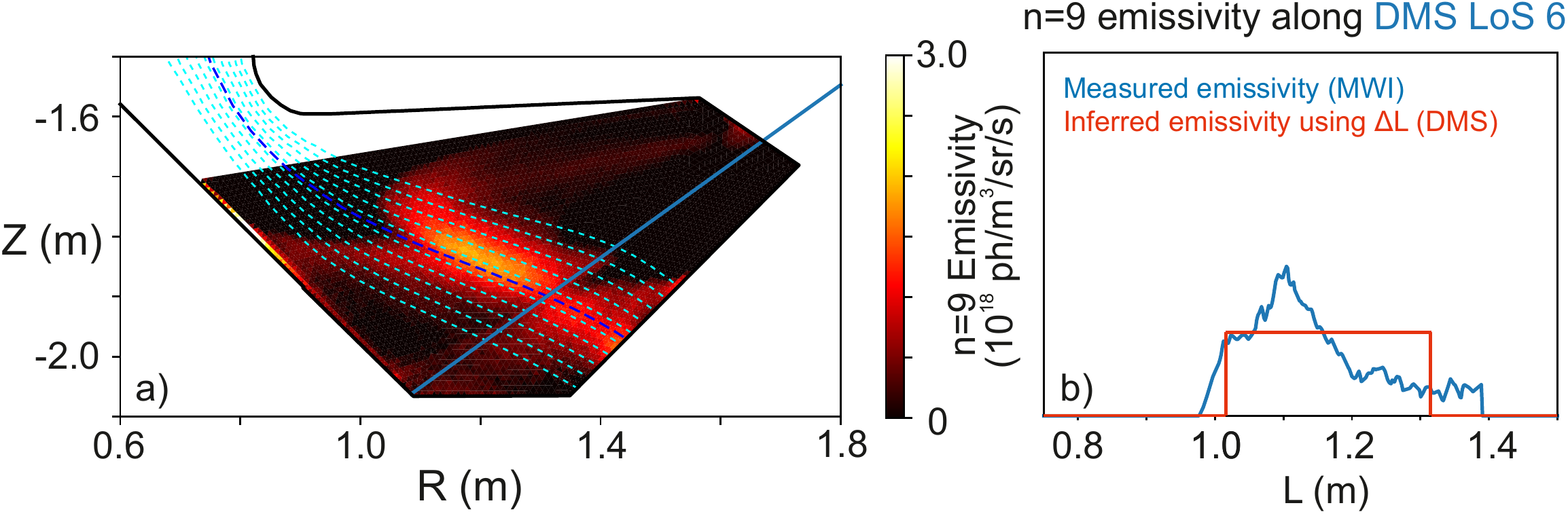}
\caption{a) Emissivity of the $n=9$ Balmer line, obtained by inverting the MWI imaging data, at 0.755 s for \# 45370 with DMS line-of-sight (LoS) 6 indicated. b) $n=9$ Balmer line emissivity profile along the DMS line-of-sight 6, with an overlaid emissivity profile assuming a constant emissivity along the inferred $\Delta L$ and using the DMS measured $n=9$ brightness.}
\label{fig:MWI_n9_inversion_pathlength}
\end{figure}

Assuming the high-n Balmer line emission brightness is fully dominated by EIR, the inferred $T_e$ and $n_e$ from the high-n Balmer line fit can be used to estimate the emission path-length \cite{Verhaegh2023}: $\Delta L = \frac{B_{n\geq9}}{n_e^2 \sum_{n\geq9} PEC_{n}^{EIR} (n_e, T_e)}$. For the fit shown in figure \ref{fig:Highn_fit_Te}, the obtained $\Delta L$ using the new ADAS data was found to be $\Delta L = 0.30$ $(0.27-0.35)$ m ($T_e = 0.09$ $(0.08-0.1)$ eV) for a deeply detached discharge with high-n $(n\geq9)$ coverage of the Balmer lines. Assuming $T_e=0.2$ eV instead (default ADAS limit), the obtained path lengths would be longer than the DMS line of sight ($\Delta L = 1.8$ $(1.6 - 2.0)$ m assuming the same $n_e$). The path lengths can also be estimated experimentally by inverting the MWI observations \cite{Feng2017,Wijkamp2023}, which was shown for a specific discharge in figure \ref{fig:MWI_n9_inversion_pathlength} where it is compared to the DMS emissivity profile using the measured DMS brightness and the inferred $\Delta L$  obtained using the new ADAS data, showing fair agreement. The measured brightness and that obtained by integrating the DMS chord along the MWI inversion agrees within 2.5\%. 

%\footnote{The high-n Balmer line fit, when using the default ADAS data, leads to higher $n_e$ estimates than the Boltzmann/new ADAS data fits. Using those higher densities, the path-length estimates using the default ADAS data become $\Delta L = 0.85 (0.71 - 0.94)$ m. However, the wings of the high-n Balmer lines are not well fitted using the default ADAS data at those higher densities}.

\section{Inferring ion flow profiles using particle balance}
\label{ch:plasmaflows_calc}

Mass conservation (equation B.1) can be used to estimate the flow profile qualitatively. Here, $S_{ion} (x)$ and $S_{rec} (x)$ are the volumetric ion sources and sinks (ions/$m^3$/s) and $v$ is the velocity. 

\begin{eqnarray}
\frac{d}{d x} n_e(x) v(x) &= S_{ion} (x) - S_{rec} (x) \\
A_{eff}(x) \frac{d}{d x} n_e(x) v(x) &= A_{eff}(x) S_{ion} (x) - A_{eff} S_{rec} (x) \\
v(\xi) &= \frac{\int_{up}^\xi A_{eff} (x) (S_{ion} (x) - S_{rec} (x)) dx + I_u}{A_{eff} (\xi) n_e (\xi)} 
\label{eq:FlowProfile}
\end{eqnarray}

The spectroscopic estimates of the divertor ion sources and sinks are, however, chordally integrated; and the ion target fluxes have been integrated over the outer target. To utilise mass conservation for such spatially integrated quantities, we multiply both sides of \ref{eq:FlowProfile} with an effective area $A_{eff} (x)$. In this case, $A_{eff} n_e v \mid_{target} = I_t$ and $A_{eff} n_e v \mid_{upstream} = I_u = I_i - I_r + I_t$ - equation \ref{eq:ParticleBalance}). Here, $I_i$, $I_r$ are the ion sources and ion sinks inferred spectroscopically, integrated over the entire divertor domain (figure \ref{fig:PartBal}), which can be expressed as $\int_{up}^{target} A_{eff} (x) (S_{ion} (x) - S_{rec}(x)) dx$. Performing this integral not over the entire domain, but up until some point $\xi$ allows us to obtain a qualitative measure of the effective area times the flow velocity, as indicated in equation \ref{eq:FlowProfile}. The effective area $A_{eff}$ at a position along the divertor leg, is approximated as $2 \pi R(x) \Delta L$, where $\Delta L = 0.01$ m is assumed (characteristic assumed width of the flux bundles carrying most heat, based on MAST upstream scalings mapped to the target using the poloidal flux expansion \cite{Harrison2013}) and $R(x)$ is the radius of the separatrix position at some distance with respect to the target ($x$). 

An electron density profile is also required for estimating the flow profile (equation \ref{eq:FlowProfile}) and for this, the electron density inferred through Stark broadening is used. Although the absolute uncertainty on the inferred Stark electron density is significant, this is a systematic uncertainty and the relative trends are more reliable (see \cite{Verhaegh2023}). Since the density, $A_{eff}$ and $I_t$ cannot be relied upon quantitatively, the flow velocity at the target is scaled such that it matches (at least) the sound speed at the target, according to the Bohm criteria \cite{Stangeby2000}. The target temperature was estimated spectroscopically as an average of the characteristic temperature for the EIE and EIR emission regions, weighted by the relative radiative losses of both processes. This leads to a scaling factor of $1.7-2.8$, increasing the ion flow velocity. This may suggest that the electron density near the target is overestimated by the Stark broadening analysis, which was suspected in \cite{Verhaegh2023} and would reduce the inferred flow velocity (equation \ref{eq:FlowProfile}).

\section{Collisional-radiative modelling of $H_2 (\nu)$}
\label{ch:colrad}

Hydrogen rates are employed in CRUMPET for the collisional-radiative modelling of $H_2 (\nu)$, which are listed in \autoref{tab:Reacs}. The higher isotope mass and thus lower relative velocity between $D^+$ and $D_2$ at the same ion temperature has been accounted for by employing ion isotope mass rescaling to the ion temperature dependant rates. It is assumed that the ion temperature equals the electron temperature. and that the $D^+$ density. An electron density of $10^{19} m^{-3}$ has been assumed and it is assumed that the $D^+$ density equals the electron density. In this collisional-radiative model calculation, $H_2(\nu = 0)$, $H$, $e^-$ and $H^+$ have been set up as reservoir species.

\begin{table}[h!]
\begin{tabular}{lll}
Description                              & Reaction                                                                                                                        & Data                                                  \\
Vibrational excitation  & $e^- + H_2 (\nu) \rightarrow e^- + H_2 (\nu \pm 1)$                                                                             & Eirene / 'H2VIBR'  \cite{Reiter2005} \\
$H_2$ ionisation                         & $e^- + H_2 (\nu) \rightarrow 2 e^- + H_2^+$                                                                                     & Eirene / 'H2VIBR'  \cite{Reiter2005} \\
$H_2$ dissociation                       & $e^- + H_2 (\nu) \rightarrow e^- + H + H$                                                                                        & Eirene / 'H2VIBR' \cite{Reiter2005}  \\
Molecular charge exchange                & $H^+ + H_2 (\nu) \rightarrow H_2^+ + H$                                                                                         & Ichihara \cite{Ichihara2000}         
\end{tabular}
\caption{Table showing reactions used for collisional-radiative modelling.}
\label{tab:Reacs}
\end{table}



%% References







% \end{thebibliography}

\end{document}